\documentclass[a4paper,fleqn,usenatbib]{mnras}  
\usepackage{newtxtext,newtxmath}
\usepackage[T1]{fontenc}
\usepackage{ae,aecompl}
\usepackage{graphics,graphicx}
\usepackage{amsmath}	

\def\IDCS{IDCS J1426.5+3508 }

\newcommand{\arcsecs}{$^{\prime \prime}\ $}
\newcommand{\arcmins}{$^{\prime}\ $}

\title[\IDCS thermodynamic evolution]{
Thermodynamic evolution of the $z=1.75$ galaxy cluster \IDCS
}
\author[S. Andreon et al.]{S. Andreon,$^{1}$\thanks{E-mail:stefano.andreon@inaf.it} 
C. Romero$^{2,3}$, 
F. Castagna$^{1}$,
A. Ragagnin$^{4,5}$,
M. Devlin$^3$,
S. Dicker$^3$, \and
B. Mason$^6$,
T. Mroczkowski$^7$,
C. Sarazin$^8$,
J. Sievers$^9$, and
S. Stanchfield$^3$ \\
$^1$ INAF--Osservatorio Astronomico di Brera, via Brera 28, 20121, Milano, Italy\\
$^2$ Green Bank Observatory, 155 Observatory Road,  Green Bank, WV 24944, USA \\
$^3$ Department
of Physics and Astronomy, University of Pennsylvania, 209 South 33rd Street, Philadelphia, PA, 19104, USA \\
$^4$ IFPU - Institute for Fundamental Physics of the Universe, Via Beirut 2, 34014 Trieste, Italy\\
$^5$ INAF - Osservatorio Astronomico di Trieste, via G.B. Tiepolo 11, 34143 Trieste, Italy \\
$^6$ National Radio Astronomy Observatory, 520 Edgemont Rd, Charlottesville, VA 22903\\
$^7$ European Southern Observatory, Karl-Schwarzshild-Str. 2, D-85748 Garching b. M\"{u}nchen, Germany \\
$^8$ Department of Astronomy, University of Virginia, P.O. Box 400325, Charlottesville, VA 22904, USA \\
$^9$ Department of Physics, McGill University, 3600 University Street, Montreal, QC H3A 2T8, Canada \\
}
\date{Accepted 2021 June 03. Received 2021 May 03; in original form 2020 October 09}
\pubyear{2020}

\begin{document}
\label{firstpage}
\pagerange{\pageref{firstpage}--\pageref{lastpage}}
\maketitle

\begin{abstract}
We present resolved thermodynamic profiles out to 500 kpc,
about $r_{500}$, of the $z=1.75$ galaxy cluster \IDCS  with 40 kpc resolution. 
Thanks to the combination of
Sunyaev-Zel'dovich and X-ray datasets,
\IDCS becomes the most distant cluster with
resolved thermodynamic profiles. 
These are derived assuming a non-parametric pressure profile and
a very flexible model for the electron density profile.
The shape of the pressure profile
is flatter than the
universal pressure profile.
The \IDCS temperature profile is increasing radially out to 500 kpc. 
To identify the possible future evolution of
\IDCS , we compared it with its local descendants
that numerical simulations show to be
$0.65\pm0.12$ dex
more massive.
We found no evolution
at 30 kpc, indicating a fine tuning between cooling and heating
at small radii.
At $30<r<300$ kpc, 
our observations show that
entropy and heat must be
deposited with little net gas transfer, while
at 500 kpc the gas need to be replaced
by a large amount of cold, lower entropy gas, consistent with theoretical expectation of a filamentary 
gas stream, which brings low entropy 
gas to 500 kpc and energy at even smaller radii.  
At $r \gtrsim 400$ kpc
the polytropic index takes a low value, 
which indicates 
the presence of a large amount of non-thermal pressure.
Our work also introduces a new definition of
the evolutionary rate, which uses unscaled radii, unscaled thermodynamic quantities, and 
different masses at different redshifts to compare ancestors and descendants.  It has the advantage of separating cluster evolution, 
dependence on mass, pseudo-evolution and returns a number with unique interpretation, unlike other definitions used in literature.
\end{abstract}

\maketitle

\begin{figure}
\centerline{\includegraphics[width=9truecm]{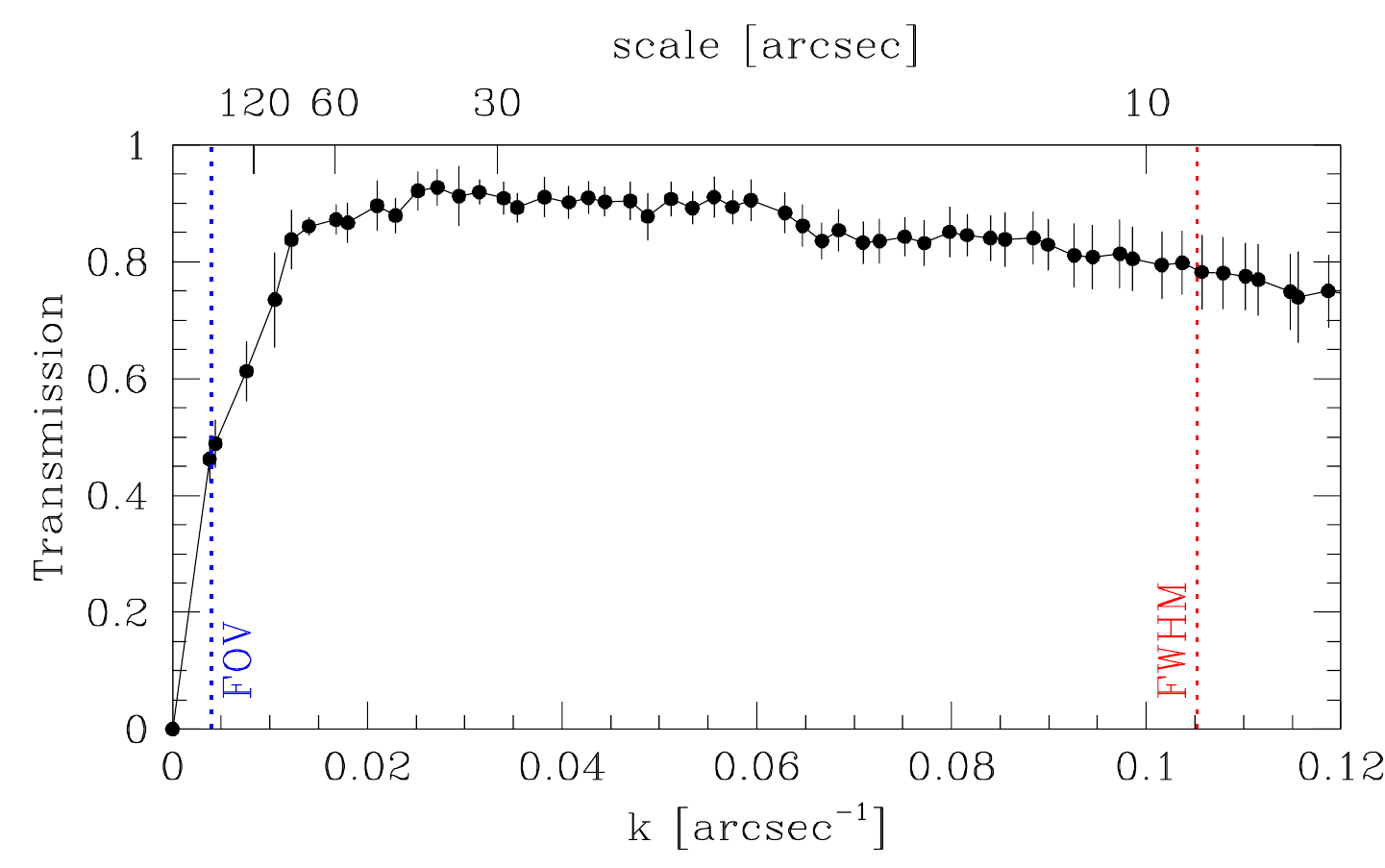}}
\caption[h]{\IDCS MUSTANG-2 transfer function. Signal at scales smaller than 1.5 arcminutes is, by large,
preserved by the MIDAS data reduction but is
increasingly reduced with increasing scales.
}
\label{fig:tf}
\end{figure}

\begin{figure*}
\centerline{
\includegraphics[width=7truecm]{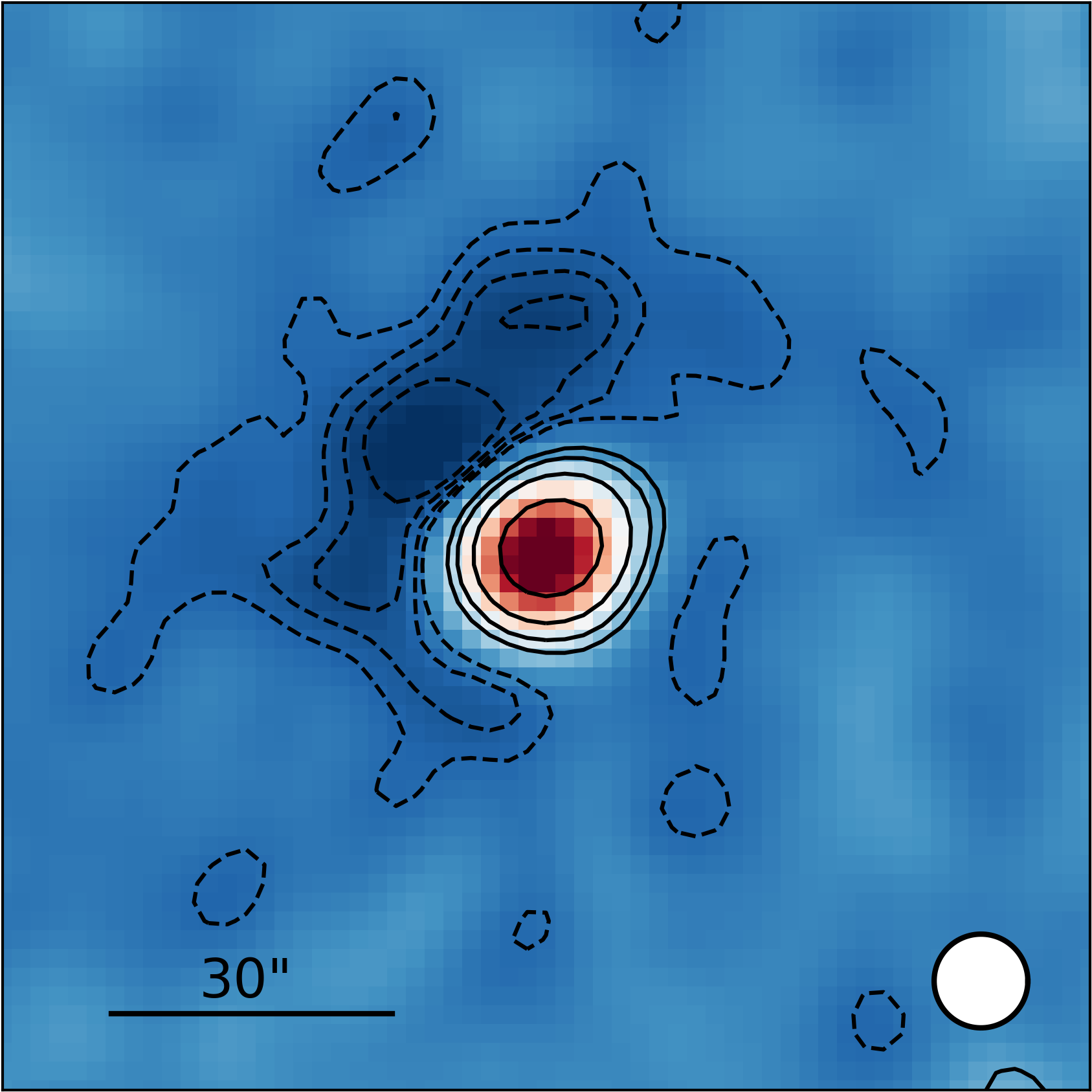}
\includegraphics[width=7truecm]{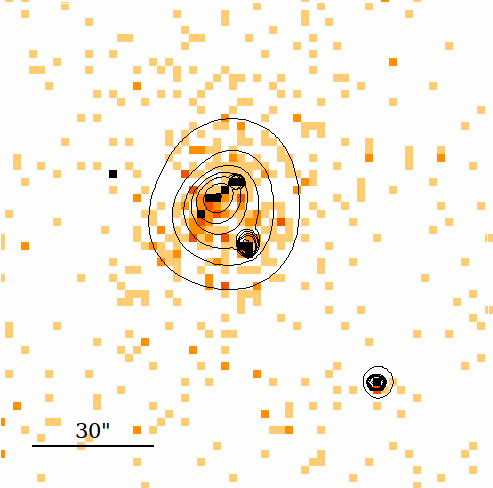}}
\caption[h]{Multi-wavelength view of \IDCS. {\it Left panel:} MUSTANG-2 SZ map; negative contours start at -2$\sigma$, with $1\sigma$ steps. Positive contours start at $3\sigma$ and increment by a factor of 2. A bright point source is obvious at SW. The MUSTANG-2 beam is indicated in the bottom right. {\it Right panel:} Adaptively smoothed Chandra [0.5-2] keV contours on the top of a binned [0.5-2] keV
Chandra image. Three bright sources are obvious; the SW one at 14" from the cluster center
is also obvious in the SZ map. 
}
\label{fig:view}
\end{figure*}
\section{Introduction}

Understanding the growth of large scale structure is one of the challenges of modern astrophysics. In massive clusters  most baryons are in the intracluster medium (ICM, Vikhlinin et al. 2006; Andreon 2010 and
references therein) and the ICM offers us the 
opportunity of determining the dynamical status of clusters, their formation and evolution (Voit 2005; Kravtsov \& Borgani 2012) via the measurement
of their thermodynamic profiles (electron density, temperature, pressure, and entropy).
While these quantities can be measured with X-ray facilities such as 
the EPIC camera on XMM-Newton (Turner et al. 2001, Struder et al. 2001) and the ACIS on Chandra (Garmire
et al. 2003), the observational
cost dramatically increases with increasingly redshift. 
Consequently,
the earliest evolutionary phases, e.g. at $z\sim2$, are observationally poorly 
known (see our Sec~3.1 for examples).

Instead, the Sunyaev-Zel'dovich (SZ; Sunyaev \& Zel'dovich 1972) signal, a spectral distortion of the 
cosmic microwave background induced by inverse Compton scatter of electrons in the hot cluster gas, is largely redshift-independent, 
making the observational cost less redshift-dependent than it is in X-ray (e.g., Mroczkowski et al. 2019). 
However, to map the ICM beyond $z\approx 0.3$, a resolution of 10-20 arcsec 
is needed (20 arcsec correspond to 170 kpc at $z\approx 1$). Early interferometers
(e.g. the SZA array\footnote{http://www.mmarray.org/}) did not resolve distant clusters,
current interferometers lose the large spatial scales associated with the
ICM emission (e.g. the Atacama Compact Array, Iguchi et al. 2009, see Di Mascolo et al. 2019, 2020). 
Large  single-dish SZ telescopes, such as the IRAM 30m and the GBT, coupled with multi-feed
receivers, are needed to sample
spatial scales between 50 kpc to 700 kpc, and beyond, outside the local Universe,
and also to identify, and later flag, point sources superimposed on the extended
ICM emission. By measuring pressure, SZ instruments reduce
the required X-ray exposures, as temperatures no longer need to be derived from X-ray spectra, which are observationally expensive. Combining SZ and X-ray data thus offers the possibility of studying the thermodynamic profiles of clusters at $z\sim2$.

Only a dozen of $z\sim1 - 1.4$ clusters, and only one beyond\footnote{SPT-CL0459-4947, that we later
consider in this work, is at 
$z=1.70-1.71$ (Mantz et al. 2020; Ghirardini et al. 2021), had an initial slightly larger
photometric redshift,
and its redshift was rounded up to $z=1.9$ (Ghirardini et al. 2021; 
McDonald et al. 2019).}, have resolved pressure profiles, and 
these shows a variety of shapes: while some  are 
flatter than the universal pressure profile (Dicker et al. 2020, Ghirardini et al. 2021),
others are consistent with it (Ruppin et al. 2020, Dicker et al. 2020) and
similar to those of the local Universe (Ghirardini et al. 2021), suggestive
of a possible evolution. We emphasize, however, that studied
samples are far from being unbiased: first,  clusters with non-universal
pressure profiles exist
even at low redshift  (Andreon et al. 2019), and, second, 
pressure-selected clusters, such
as SPT clusters studied by Ghirardini et al. (2021), favour clusters
with high pressure by definition.

In this work, we compute the thermodynamic profiles for one of the most distant clusters with a resolved
electron density profile, \IDCS at spectroscopic $z=1.75$  (Stanford et al. 2012), combining deep {\it Chandra} data with MUSTANG-2 observations. 
By observing a very distant cluster, we maximize the
redshift baseline for evolutionary studies and we probe earlier times.
\IDCS was discovered in the near-infrared as a galaxy overdensity (Stanford 
et al. 2012). 
At a fixed mass, the galaxy overdensity is independent of the ICM, which is the focus of this paper\footnote{ICM and galaxy are clearly correlated in the sense 
that more massive clusters have more of everything. However,
selection effects act when the correlation is at fixed mass (e.g., Andreon \& Berg{\'e} 2012).}.
This independence has the advantage that it guarantees that the measured quantity is not biased by the way the object
is selected, unlike samples selected by pressure.

\IDCS was later targeted and detected
with the low (2\arcmin) resolution Sunayev-Zel'dovich Array (Brodwin et al. 2012). 
Mass estimates from that work (Brodwin et al. 2012) 
involving the integrated SZ signal
and deep {\it Chandra} data
(Brodwin et al. 2016) scatter around a value of about $M_{500}\sim2.5 \ 10^{14}$ M$_\odot$ ($r_{500}\sim500$ kpc;  all values are corrected
to our adopted cosmology), making the cluster massive for its redshift.

Throughout this paper, we assume $\Omega_M=0.3$, $\Omega_\Lambda=0.7$, 
and $H_0=70$ km s$^{-1}$ Mpc$^{-1}$. 
Results of stochastic computations are given
in the form $x\pm y$, where $x$ and $y$ are 
the posterior mean and standard deviation. The latter also
corresponds to 68\% uncertainties because we only summarize
posteriors close to Gaussian in this way. All logarithms are in base 10.

\section{Observations, data reduction and analysis}

\subsection{SZ}
\label{sec:SZ}

The cluster has been observed with MUSTANG-2, a 215-element array
of feedhorn-coupled TES bolometers (Dicker et al. 2014) at the
100m Green Bank telescope (GBT). When coupled to the GBT, MUSTANG-2
has 10"  FWHM resolution and an instantaneous field of view of 265 arcsec.

\IDCS was observed with
Lissajous daisy scans, typically with a 2.5 arcmin radius. The cluster
has been observed for a total on-source integration time of 5.0 hours
in April, June, and December 2018. Data
are calibrated using preferentially Solar systems objects, but also 
ALMA calibrators (Fomalont et al. 2014, van Kempen et al. 2014).
Our SZ maps are calibrated to Rayleigh-Jeans brightness temperature ($K_{\text{RJ}}$); 
throughout this paper, brightness units of $K$ are thus $K_{\text{RJ}}$ unless otherwise noted.
The final map has a RMS noise of 20 $\mu K$, when smoothed to the beam
resolution, within the central 2 arcmin radius. A beam profile specific to
this cluster observations is created by stacking all secondary (point-source)
calibrators used during these observations (see Ginsburg et al 2020).

Data are reduced following the traditional approach (named MIDAS in
Romero et al. 2020, to which we refer for details): time ordered data from unresponsive detector or
from bad scans and glitches are flagged, remaining ones are filtered to subtract
atmospheric and electronic signal. This filtering also removes
some (cluster) signal, which needs to be accounted for during the analysis
using the transfer function, shown in Fig.~\ref{fig:tf}. The transfer function is
close to 0.8-0.9 up to scales of 1.5 arcmin, then drops to lower values for larger scales.

The resulting map is shown in Fig.~\ref{fig:view}. At a glance, the negative (SZ) signal has a
pronounced elongation in the NW-SE direction; however, this is largely the effect of
a bright (positive) source sitting 14" SW of the cluster, 
quite obvious in the map, that disappears in the point-source subtracted map. 
Because of the point-source brightness, we flagged and removed from 
further cluster analysis pixels
inside a 20" diameter circle
centered on the source. To account for the point source
flux outside the aperture, we used  the average beam measured from calibration observations interleaved
with the cluster, regularized with a sum of two Gaussian, and scaled in
intensity to match the point source. As a further precaution
for pixels outside of the aperture the  
errors are boosted by 5\% of
the subtracted flux to account for uncertainties in point
source subtraction. This extra uncertainty has a minimal impact in the error budget.
5\% is a conservative estimate of the variation of individual beam observations.
To compute the scaling factor of the beam to the point source, we fit the beam profile to pixels inside
the aperture, assuming a
constant level 
for the cluster plus background
inside the small source aperture. 

\begin{figure}
\centerline{\includegraphics[width=9truecm]{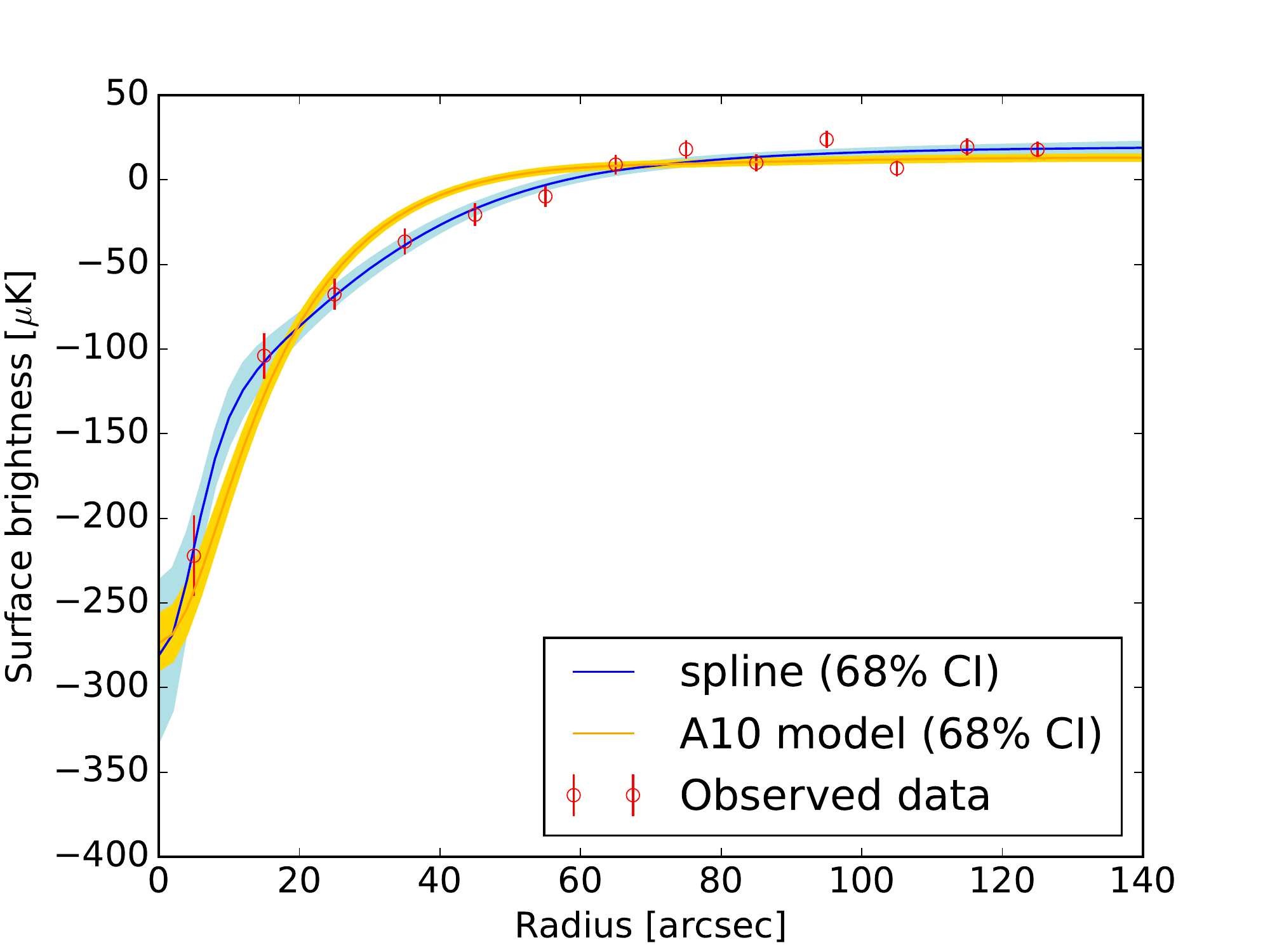}}
\caption[h]{\IDCS surface brightness profile (points with error bars) and 68\% uncertainties on the fitted model. The adopted spline for the pressure profile is able to well describe the observed data, whereas a pressure
profile with parameters fixed to A10 does not. In the latter case, model errors are meaningless.
}
\label{fig:surf_bright}
\end{figure}

Figure~\ref{fig:surf_bright} shows the cluster radial surface brightness profile, 
extracted with 10\arcsecs radial annuli.
Our forward model pressure profile, later fitted to these data, adopts the same radial bins, thus making the fitting results independent of the adopted annular width.
The profile levels off to $\approx10$ $\mu$K, 
and therefore our analysis requires one additional parameter, that we name the pedestal, to model
this non-zero level. The non-zero pedestal
originates from the data reduction, which forces the map to have zero mean.

\begin{figure}
\centerline{\includegraphics[width=9truecm]{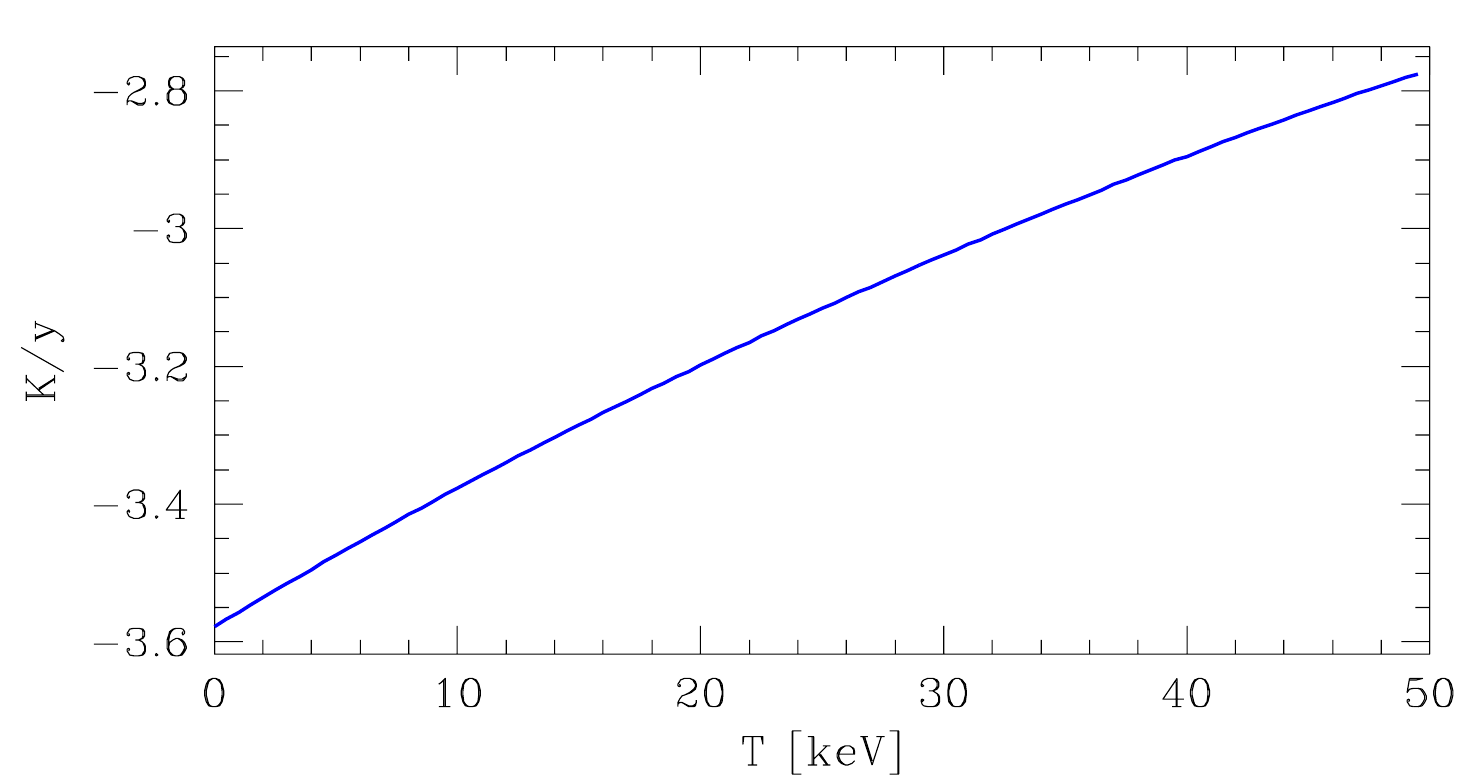}}
\caption[h]{Cluster temperature dependence of the conversion from Compton y to Kelvin.} 
\label{fig:conversion}
\end{figure}

The analysis of SZ data also requires the point spread function (PSF), measured on the beam
image, and the conversion from $K$ to Compton $y$, computed
using relativistic corrections from Itoh et al. (2004) as shown in Fig.~\ref{fig:conversion}.
The conversion is  described by a second order polynomial, $-9.126 \ 10^{-5} T^2 + 0.02067 T - 3.575$, 
with better than 0.1\% precision. 
The conversions between brightness temperature ($K_{RJ}$) and Compton $y$
accounts for conversion to $K_\text{CMB}$ and the bandpass for MUSTANG-2, which includes optical filters and Ruze efficiency of the primary dish on the GBT.

The three dimensional pressure profile is derived fitting the SZ data with a modified version of \texttt{PreProFit} (Castagna \& Andreon 2019), accounting for the transfer function, PSF, and pedestal level.  Rather than fit the shape of the pressure profile to a pre-determined model (e.g., a generalized Navarro, Frenk, \& White (1997) model with parameters fixed to the Arnaud et al. (2010) values), we allow the shape of the pressure profile to vary almost arbitrarily.   However, for our subsequent analysis, we require a pressure profile that is continuous and doubly-differentiable.  Thus, we fit the observed SZ data with a cubic spline in log-log space with knots at radii of $r=5,15,30,$ and $60$ arcsec.  Our model then has 5 variables:  the pressures at these four points, $P_0,P_1,P_2,$ and $P_3$, and the pedestal value of the SZ surface brightness. By defining
the spline in log quantities (log pressure vs log radius), we naturally exclude non-physical (negative) values
of pressure and radius and we can approximate a large variety of profiles, and their
derivatives, including the commonly pressure profile parametrised as in Nagai et al.
(2007).  In
our observational setting,
the adopted choice is advantageous compared to the Nagai et al.
(2007) parametrisation because in the latter the degeneracy between pedestal parameter and the scale radius 
propagates to the three remaining parameters.

Prior to jointly fitting SZ and X-ray data (Section~\ref{sec:SZpX}), we perform a SZ-only analysis which assumes spherical symmetry and $kT=5$ keV. We also assume 
uniform priors, zeroed for unphysical values (e.g., pressure cannot be negative).
In particular, since the total mass of the cluster is finite, the logarithmic slope
of pressure should be steeper than $-4$ at large radii (Romero et al. 2018). We therefore
adopted a logarithmic slope $<-2$ at $r=1$ Mpc ($\sim 2 r_{500}$) as prior. This prior
could alternatively be expressed as a maximal value for the pedestal level.
The imposition of this prior is primarily necessary due to the removal of cluster signal at these larger
scales (the reduction itself is agnostic as to the true background level).

\begin{figure}
\centerline{\includegraphics[width=9truecm]{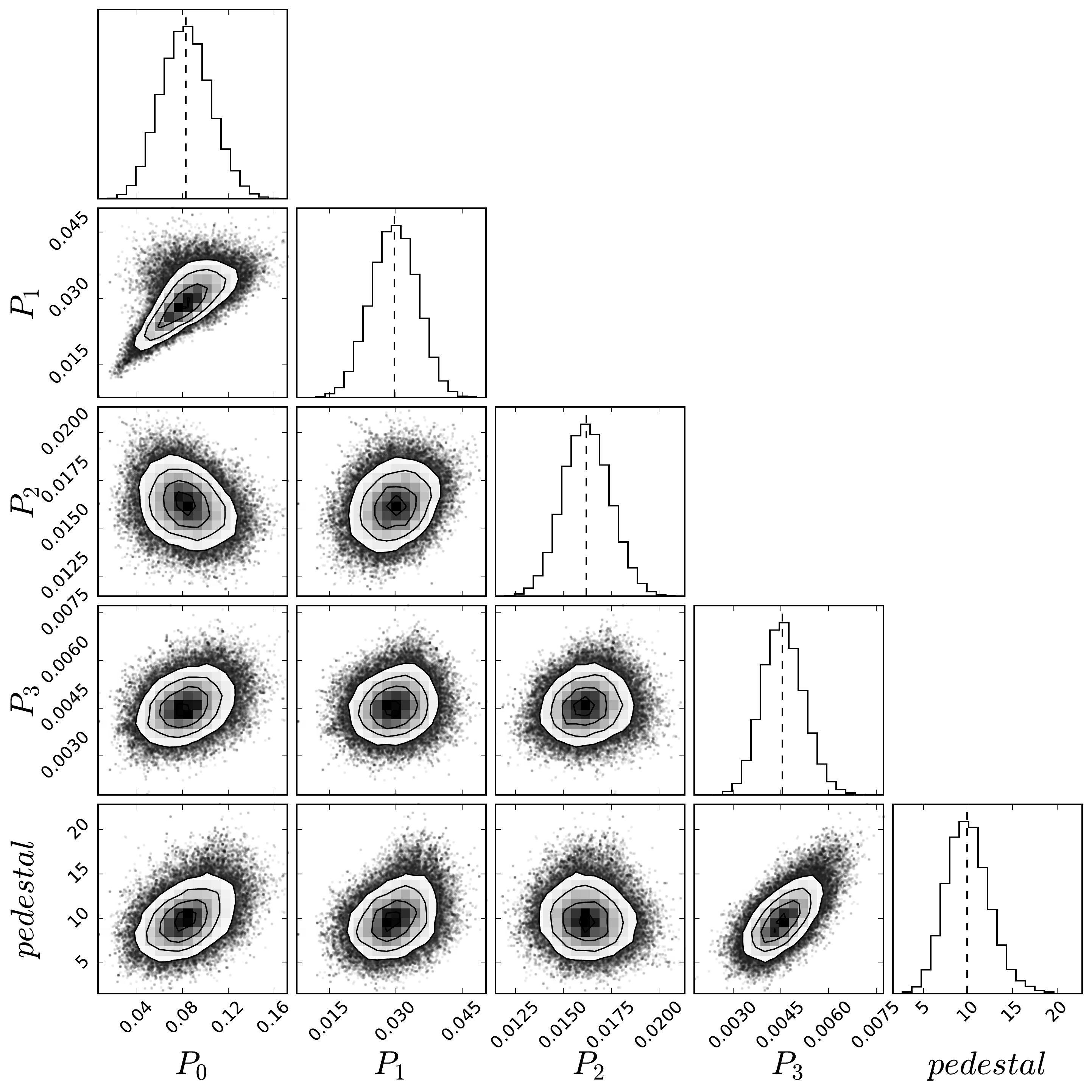}}
\caption[h]{Joint and marginal probability contours for our SZ-only fit. Units: keV cm$^{-3}$ for
pressure parameters and $\mu K$ for pedestal. 
}
\label{fig:cornerplot_sz}
\end{figure}

\begin{figure}
\centerline{\includegraphics[width=9truecm]{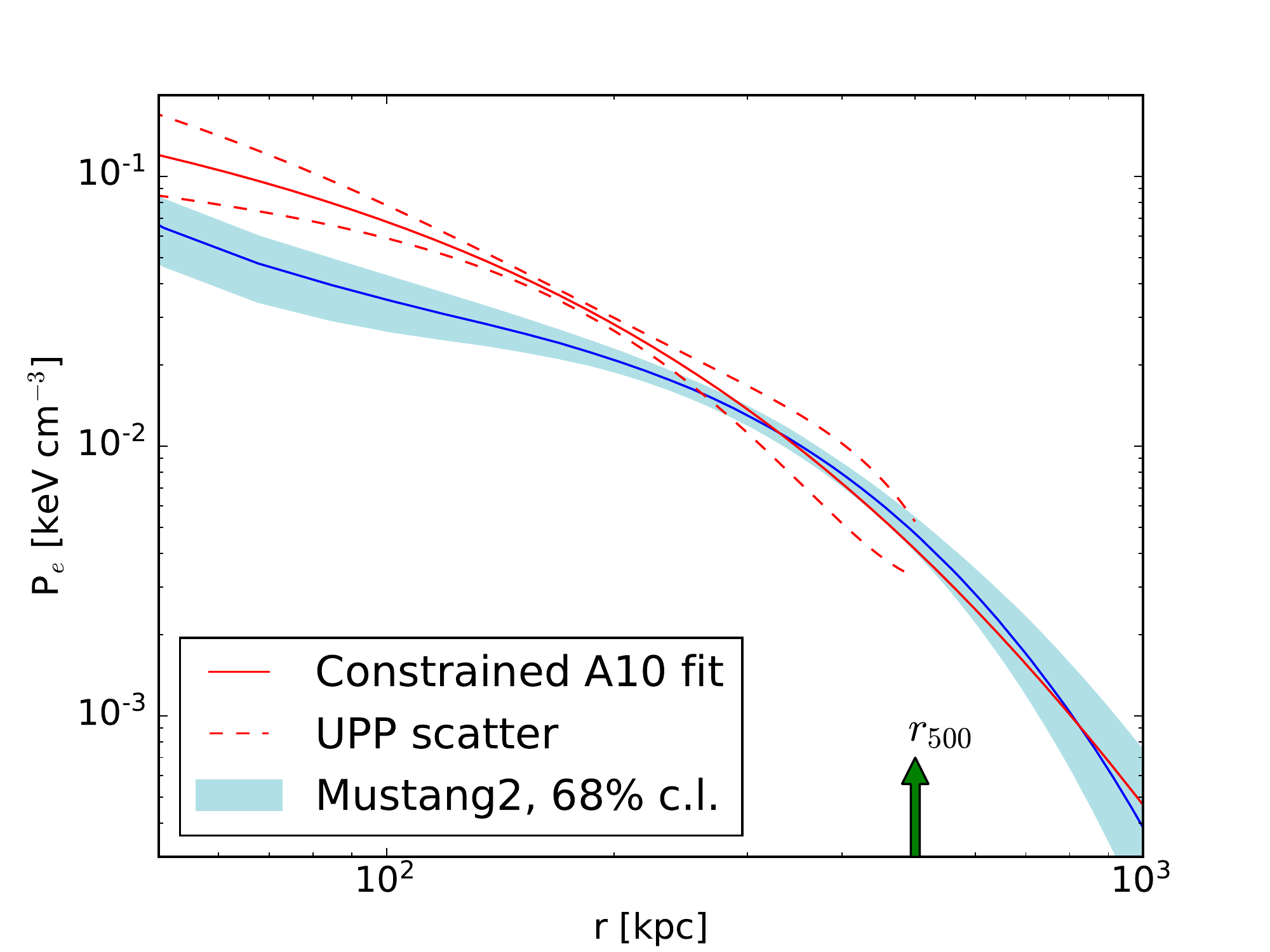}}
\caption[h]{Observed MUSTANG-2 \IDCS pressure profile with 68\% uncertainties (solid curve and shading) 
versus a constrained universal profile 
fit to low resolution SZ interferometric data. Note that the displayed radial range includes radii poorly
sampled by CARMA data  ($r\ll 1$ arcmin, about 500 kpc), or MUSTANG-2 ($r\gtrsim 1$ arcmin). There is a good agreement on scale
where interferometric data are most sensitive, and notable differences (25\% reduction) at smaller radii. The Arnaud et al. (2010) scatter around the universal profile is also indicated.}
\label{fig:pressure}
\end{figure}

We find $P_i=0.083\pm0.022, 0.029\pm0.005, 0.016\pm 0.001, 0.0045\pm0.0006$ keV cm$^{-3}$, 
and a pedestal level of $10\pm3$ $\mu$K with little covariance,
illustrated in Figure~\ref{fig:cornerplot_sz}. We found a $\chi^2$ of $12.0$ for 8
degrees of freedom. Figure~\ref{fig:surf_bright} shows the best-fit model, with the 68\% uncertainty, on top of
the observed data. Figure~\ref{fig:pressure} shows the SZ-derived pressure profile as a
blue line with 68\% uncertainty as the shaded area.
This is the first $z>1.45$ cluster 
(Dicker et al. 2020) with a resolved SZ-based pressure profile. The pressure profile is
accurately determined out to $r_{500}$ at least, where the cluster SZ brightness is $3\sigma$
away from the pedestal when measured in a narrow annulus 10\arcsecs wide (i.e. discarding 
information at adjacent radii).

The fit adopts the X-ray centre (see next section), and accounts for a $\sim4$ arcsec astrometric offset with 
the Chandra image, the latter computed using the point source. The SZ center is 5\arcsecs away from the
X-ray center, in the direction of the point source, improving upon
the 30\arcsecs offset (and uncertainty) found on the old CARMA data.
Fit results are robust to the limiting
integration radius along the line of sight, taken to be 2.5 Mpc. We note that integration to larger radii, e.g. 5 Mpc, or
smaller one, e.g. 1 Mpc, does not appreciably change the results.

Many previous SZ analyses of high redshift clusters lacked sufficient angular resolution to allow detailed pressure profiles to be derived. For example, 
CARMA observations of XLSS122 at $z_{phot}\sim1.9$ (Mantz et al. 2014, 2018),
or of \IDCS itself by Brodwin et al. (2012)  
do not resolve the cluster and thus require a pressure profile shape be assumed.
The same assumption is required for the analysis of many
interferometric observations of clusters at lower redshift
(e.g., the AMI Consortium et al. 2013, Rumsey et al. 2016, Brodwin et al. 2015), for
the detection of Planck clusters (e.g., Planck Collaboration, 2013), for estimating masses from ACT data (e.g. Hilton et al. 2018), and for analyses of
interferometric observations only sampling the smallest scales, such
as with the Atacama Compact Array (Di Mascolo et al. 2020).
The resolved pressure profile of \IDCS allows us to test  
this common assumption on the profile shape at very high redshift.
Figure~\ref{fig:pressure} compares the \IDCS pressure profile with the fitted profile from Brodwin et al. (2012). To fit their low (2') resolution interferometric data, Brodwin et al. (2012) assume the universal pressure profile as determined by Arnaud et al. (2010; hereafter, A10) with the additional constraint that the cluster has to lay exactly (with no scatter) on the $Y_{500}-M_{500}$ relation of Andersson et al. (2010), yielding a fit with effectively one degree of freedom.
There is a good agreement on arcminute scales (1 arcmin is $\approx 500$ kpc)
where CARMA interferometric data are most sensitive, and also notable differences (25\% reduction)
at smaller radii,  which are not sampled well by CARMA. This illustrate the value of high angular
resolution SZ data, particularly if the field of view is sufficiently large to cover the entire cluster.
Fig.~\ref{fig:surf_bright} quantifies the amplitude of
the discrepancy by fitting the surface brightness profile
with one free parameter, as done in many papers (plus free pedestal, requested by our data).
More precisely, we hold $r_p$ to 500 kpc and
the other parameters to the Arnaud et al. (2010) value.
This profile is clearly rejected by our data ($\chi^2=36.4$ for 11 degree of freedom).
Some other clusters at lower redshift, $z_{med} = 1.2$, show  deviations from the universal
pressure profile as those we observe for \IDCS, either a lower central pressure or an excess at large radii
(Dicker et al. 2020). In the local Universe, the cluster CL2015 (Andreon et al. 2019) 
shows an even more remarkable flat pressure profile, a factor of 3 under-pressured compared
to the universal profile in the inner 500 kpc.

\begin{figure*}
\centerline{\includegraphics[trim={20 205 50 30}, clip,height=4truecm]{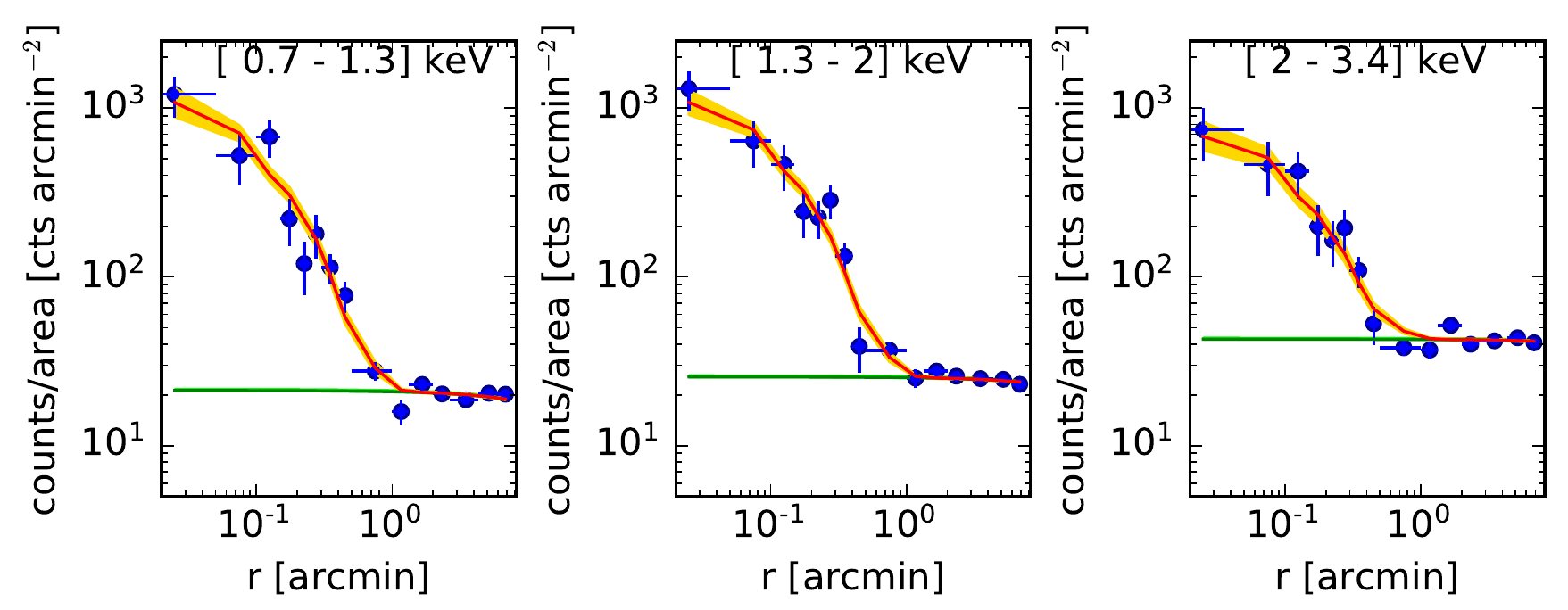}
\includegraphics[trim={20 5 200 230}, clip,height=4truecm]{fitwithmodbin.pdf}
}
\caption[h]{\IDCS surface brightness profile (points with error bars) in the X-ray bands, after coadding
the two pointings and adjacent energy bands, with 68\% uncertainties
on the fitted model (red line and yellow shading). 
The green line with lime shading (barely visible) is the background radial
profile and its 68\% uncertainty. The analysis
uses Poisson errors, separated pointings and bands, not the plotted values and 
errorbars. The assumed X-ray cluster model (red line)
captures the observed X-ray profiles at all energies.
}
\label{fig:x-ray}
\end{figure*}

The mismatch between \IDCS and the universal pressure profile 
is perhaps negligible for the derivation of the integrated Comptonization based
on observations sampling large scales, because the 
integral is dominated by the signal at large radii\footnote{In the
case of \IDCS,  the spherically averaged dimensionless Comptonization within $r_{500}$ 
is $7.0\pm0.5 \ 10^{-12}$ vs $7.9\pm3.2 \ 10^{-12}$ in Brodwin et al. (2012).}. 
However, the mismatch becomes 
crucial when the shortest spatial scales are used to recover the signal at larger scales, as 
already noted by Di Mascolo et al. (2020) for Atacama Compact Array observations.
This highlights the importance
of sampling both small and large spatial scales: Di Mascolo et al. (2020) hypothesize departures from the  universal
pressure profile to explain differences between integrated pressures measured
with CARMA and Atacama Compact Array. Hilton et al. (2018) make the
same hypothesis to explain the difference between masses (actually, integrated pressure) estimated from
Planck and ACT, which are sensitive to different spatial scales.
This work (Figs.~\ref{fig:pressure} and \ref{fig:surf_bright}) and Dicker et al. (2020) show these departures thanks to a resolved
pressure profiles measured with an instrument sensitive to both
small and large scales.
The nature of these departures are characterized below
with complementary probes of the ICM, such as those accessible from X-ray observations.

\subsection{X-ray}
\label{sec:X}

The \emph{Chandra} ACIS-I X-ray data (obs ID 15168, 16321 with total exposure time of 100 ks; PI: Brodwin) were reduced following the standard procedures (e.g., Castagna \& Andreon  2020).
We briefly review the procedures here. We first removed background flares. 
Subsequently, we detect point sources using a wavelet detection algorithm and we masked them. Events (photons)
are partitioned into ten bands ([0.7-1.0], [1.0-1.3], [1.3-1.6], [1.6-2.0], [2.0-2.7], [2.7-3.4], [3.4-3.8], [3.8-4.3], [4.3-5.0], and [5.0-7.0] keV). To calculate the effective exposure time, we computed
energy-dependent exposure maps accounting for vignetting,
dithering, gaps, CCD defects, and flagged pixels.  To measure the radial profiles, we then adopted 
circular annuli with an increasing width with radius
to reduce the impact of the decreasing intensity of the cluster. 
The minimal width 
is taken to be 3 arcsec, larger than the Chandra PSF.
We only considered annuli where the exposure time is larger than 50\% of the on-axis
exposure time and included in the field of view by more than half of its area. 
The cluster centre is iteratively
computed as the centroid of X-ray emission within the inner 20 kpc. 
The ancillary response and redistribution matrix files are derived using CIAO tools
(with CALDB 4.8.3).

Proper background modelling is important for spectral analysis,
especially in regions of low surface brightness.
We use blank field images using CIAO \textsc{blanksky} (Fruscione et al. 2006), 
normalized to
the count rate in the hard band [9-13] keV to derive the background surface brightness in
the ten bands accounting for exposure time variations, dithering, vignetting, CCD defects,
gaps, and flagged pixels. Background fields are mostly used to measure the spatial
shape of the background in the various bands, or, equivalently, the spatial
dependence of the background spectrum\footnote{The background has a spatial-dependent
shape because it is formed by a vignetted and unvignetted part, and vignetting is
energy-dependent.}. The background normalization is constrained in the fit by the counts in the
cluster direction at large clustercentric distances. 

Cluster observations are taken with two different orientations and pointings, and, as a consequence
CCD gaps and sky coverage will differ between the two pointings. Furthermore, exposure
time and background level differ between the pointings. 
These pointings are treated as independent datasets and are used jointly in our fits.
In essence, we have four data cubes (two pointings, each with 
a cluster and a background cube) with ten
bins in energy and 15 radial bins, that we plot as radial profiles in different bands, 
shown in Fig.~\ref{fig:x-ray} (points and green line) after co-adding the two pointings and adjacent
energy bands for illustrative purposes only. Each cube may also be alternatively seen as
ten-point spectra taken at 15 different clustercentric distances.

There are 520 net photons within 4\arcmins of the cluster center in
the [0.7-7] keV band.  Given the low number of net cluster photons
over a 100 ks long background, faint morphological
features, as surface brightness edges or cavities, cannot be detected. 
The X-ray peak is clearly displaced
with respect to the brightest central galaxies by $\approx 15$ kpc, 
as already remarked by Brodwin et al. (2016) using the very same data,
indicative of a recent merger or gas sloshing.

In the X-ray only analysis of this section, SZ data are ignored and thermodynamic profiles were derived using \texttt{MBProj2} (Sanders et al. 2018), a Bayesian
forward-modeling projection code that fits the data cube accounting for the background. 
As with other approaches, \texttt{MBProj2} makes the usual assumptions about cluster 
sphericity, clumping, and, when requested, hydrostatic equilibrium. However,
it goes beyond the limitations of previous approaches, which 
apply arbitrary regularization kernels to deal with noise, ignore 
temperature gradients when deriving the electron density profile, 
or ignore the spectral variation of
the background, which represents the large majority of published papers, including
previous X-ray analyses of \IDCS.

We fit the data twice to understand the impact of assumptions on the derived
thermodynamic profiles.  In the first 
descriptive analysis, we opted for an approach
similar to Vikhlinin et al. (2006) and we  modeled the temperature
and electron density profiles with flexible functions constrained by the data. This is a 
fit aimed at simply describing the observations, with models introduced to impose 
regularity and smoothness. 
In particular, the \texttt{MBProj2} code models the electron
density as a modified single-$\beta$ profile following Vikhlinin et al. (2006):
\begin{equation}
  n_\mathrm{e}^2 = n_0^2
  \frac{(r/r_\mathrm{c})^{-\alpha}}{(1+r^2 / r_\mathrm{c}^2)^{3\beta-\alpha/2}}
  \frac{1}{(1 + r^\gamma / r_\mathrm{s}^\gamma)^{\epsilon / \gamma}}.
\label{eq:dens}  
\end{equation}

Similarly to Vikhlinin et al. (2006), the temperature profile is given by 
the product of a broken power law with three slopes and
a term introduced to model the temperature decline in the
core region:
\begin{equation}
\rm{T} = \textrm{T}_0\frac{((r/r'_\mathrm{c})^{a_\mathrm{cool}}+(\textrm{T}_\mathrm{min}/\textrm{T}_0))}
         {(1+(r/r'_\mathrm{c})^{a_\mathrm{cool}})}\frac{(r/r_\mathrm{t})^{-a}}{(1+(r/r_\mathrm{t})^b)^{c/b}}.
\end{equation}
The other
thermodynamic profiles are derived from the ideal gas law. 
Following Vikhinin et al. (2006), we fix $\gamma=3$. As in some of Vikhinin et al. (2006) fits, we adopt $\epsilon=5$, 
and following Sanders et al. (2018) 
we use weak priors for the remaining six free electron density parameters.
Following McDonald  et al. (2014), we
fix the inner slope to $a=0$ and the shape parameter of the inner region to $a_{cool}=2$,
and we adopt weak priors for the remaining ones.

In the second fit, we adopt instead a physical model for the cluster: we assume
a Navarro, Frenk, \& White (1997, NFW)
mass profile (for the dark matter only) and  hydrostatic equilibrium, which makes explicit temperature 
profile modeling unnecessary. 
In this case, \texttt{MBProj2} computes
the pressure profile given the NFW mass profile 
and derives the other thermodynamic profiles
combining it with the electron density profile. As in our descriptive analysis, 
parameters are determined by fitting the data.

In both fits, metallicity is a free parameter, 
absorption was fixed at the Galactic $N_H$ value in the direction of the cluster
from Kalberla et al. (2005),
and the results are marginalized over a further background scaling parameter to account
for systematics (differences in background level between the cluster and
control fields). 
The model is integrated on the same energy 
and radial bins as the observations, so that the results do not depend on the
binning choice. 

Figure~\ref{fig:x-ray} illustrates how well the physical model (red line with 68\% uncertainty shaded) 
fit the data after
co-addition and band-merging  
for illustrative purposes
only. The fit uses instead the four data cubes.

All fit parameters show
well-behaved posterior distributions with little or well-known covariances
(e.g. $\beta-r_c$). The backscale parameter, which measures the amplitude of a potential (multiplicative) offset between the X-ray backgrounds
in lines of sight of the cluster and background fields, is almost exactly one and extremely well determined, $1.006\pm 0.006$.
The gas metallicity is large, albeit with large errors: $Z_{Fe}=0.8\pm0.5$ Solar.
We verified that thermodynamic profiles are minimally affected by fixing metallicity at $0.3$ Solar.

\begin{figure}
\centerline{\includegraphics[width=9truecm]{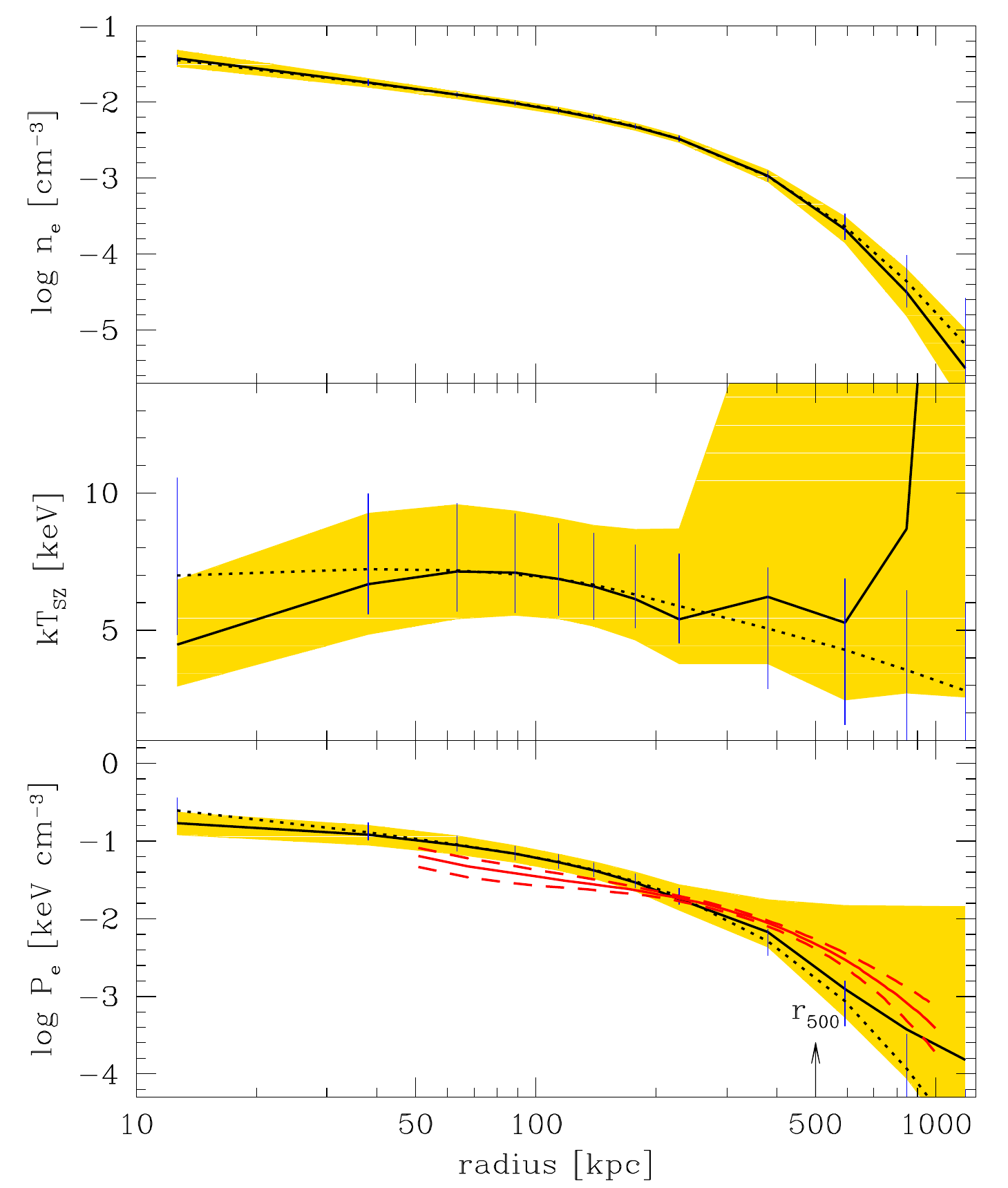}}
\caption[h]{\IDCS electron density (top panel), temperature (middle panel), and pressure (bottom) profiles (mean value and 68\% uncertainties) from the X-ray only analysis (the physical
fit is indicated by a solid line and yellow shading, while the descriptive
fit by a dashed line and error bars; in the upper panel these are almost
identical).
The red solid line with the dashed corridor in the bottom panel indicates the 
results of the SZ analysis only (mean value and 68\% uncertainties). The SZ pressure profile is extrapolated
beyond the outermost knot at $r\sim 500$ kpc.}
\label{fig:thermo_profs_x}
\end{figure}

The electron density profile (upper panel of Fig.~\ref{fig:thermo_profs_x})
is robustly determined and independent of the assumed model 
because it is primarily derived from the deprojected surface brightness profile that we can trace to 
2 arcmin ($\sim 1$ Mpc, $\approx 2 r_{500}$).  

Information on the X-ray temperature profile is encoded in the ratio 
of the cluster brightness in different energy bands, and temperature gradients are 
derived from radial variations in these ratios. 
X-ray data alone poorly constrain the \IDCS  temperature profile (central panel of Fig.~\ref{fig:thermo_profs_x})
and even less so its radial gradient. The X-ray based pressure profile shares 
limitations with the temperature profile because the pressure
is derived from (the physical fit), or constrained by (the descriptive fit), the temperature. 
Indeed, the bottom panel of Fig.~\ref{fig:thermo_profs_x} shows the two X-ray derivations of
pressure have large 68\% uncertainties that, furthermore, depend on the type of fit
at $r>200$ kpc (bars vs shading). At smaller radii, $r<200$ kpc, the X-ray derived pressure has 
errors of $\pm0.1$ dex.

In contrast, SZ data offer a more direct measurement of pressure and 
have typically negligible errors, $\sim 0.03$ dex, at radii $r<500$  kpc and are
minimally affected by PSF at $r>200$ kpc
(Fig.~\ref{fig:thermo_profs_x}, bottom panel, dashed lines).  
The constraints on thermodynamic profiles from X-rays extend to smaller radii because of the better resolution of Chandra compared to MUSTANG-2. 
X-ray and SZ pressure profiles are already consistent
at about $\sim2\sigma$ under the simplifying assumptions of this comparison:
a perfectly known 5 keV isothermal cluster (assumed in the SZ-only analysis), perfect instrument calibrations, and 
identity between the gas-mass weighted (SZ) temperature $T_{SZ}$ and the X-ray temperature $T_X$.
Nevertheless,
the SZ-derived pressure profile (solid red line) is flatter than X-ray-derived pressure profile 
(solid and dotted blue lines). Since the latter comes from a nearly constant temperature (yet very uncertain),
the flatter SZ-derived pressure profile implies that the more precisely determined 
temperature profile derived combining the
SZ-derived pressure profile and X-ray derived electron density will be radially increasing,
as quantified in the next section.

\subsection{SZ+X-ray analysis}
\label{sec:SZpX}

We now combine the SZ and X-ray data to make a fully joint fit, in which, for example
the radial-dependent SZ conversion is estimated at a temperature given by, and therefore
consistent with, the ratio of pressure and electron density profiles. We model the profile of electron density with
the Vikhlinin et al. (2006) model (Eq.~\ref{eq:dens}), and the profile of pressure with a cubic
spline, adding to our previous
analyses: a) 
one more parameter to account for a possible SZ calibration systematics (a prior with
$\sigma=10\%$ centred on 1.00; Romero et al. 2020; Dicker et al. 2020); 
b) we require consistency between the derived temperature profile and the
temperature assumed in the SZ conversion coefficient used to derive it; and c)
we allow the ratio between the X-ray and SZ temperatures to differ from 1 to account for
Chandra temperature systematics (e.g., Schellenberger et al. 2015).
In the fit, we do not assume hydrostatic equilibrium because of the
likely radially increasing temperature profile that hints at the presence of possible
disturbances. 

The fit is performed with an updated version of \texttt{JoXSZ} (Castagna \& Andreon 2020), which properly combines the 
SZ and X-ray analysis above.  Our fitting model has four nuisance parameters, over which we marginalize,
to account for data systematics: the backscale, to allow the X-ray background to deviate from the average in the cluster line of sight,
the pedestal, to compensate for the forced zero-mean SZ map, the uncertainty
on the SZ calibration, plus the parameter on the ratio of  X-ray and SZ temperature to allow both Chandra calibration systematics and intrinsic
differences between these two quantities. Furthermore, our analysis also uses the exact expression of the 
likelihood, which is a Poisson mixture model for the X-ray data (see Andreon et al. 2008; Castagna \& Andreon 2020). Finally,
since our fitting model is a fully forward one, all uncertainties are propagated to all quantities, for example
the uncertainty on the T profiles induces an uncertainty on the X-ray and SZ conversion factors (which are in passing radial-dependent) propagated to the derived T profile and the other thermodynamic quantities.

\begin{figure*}
\centerline{\includegraphics[width=18truecm]{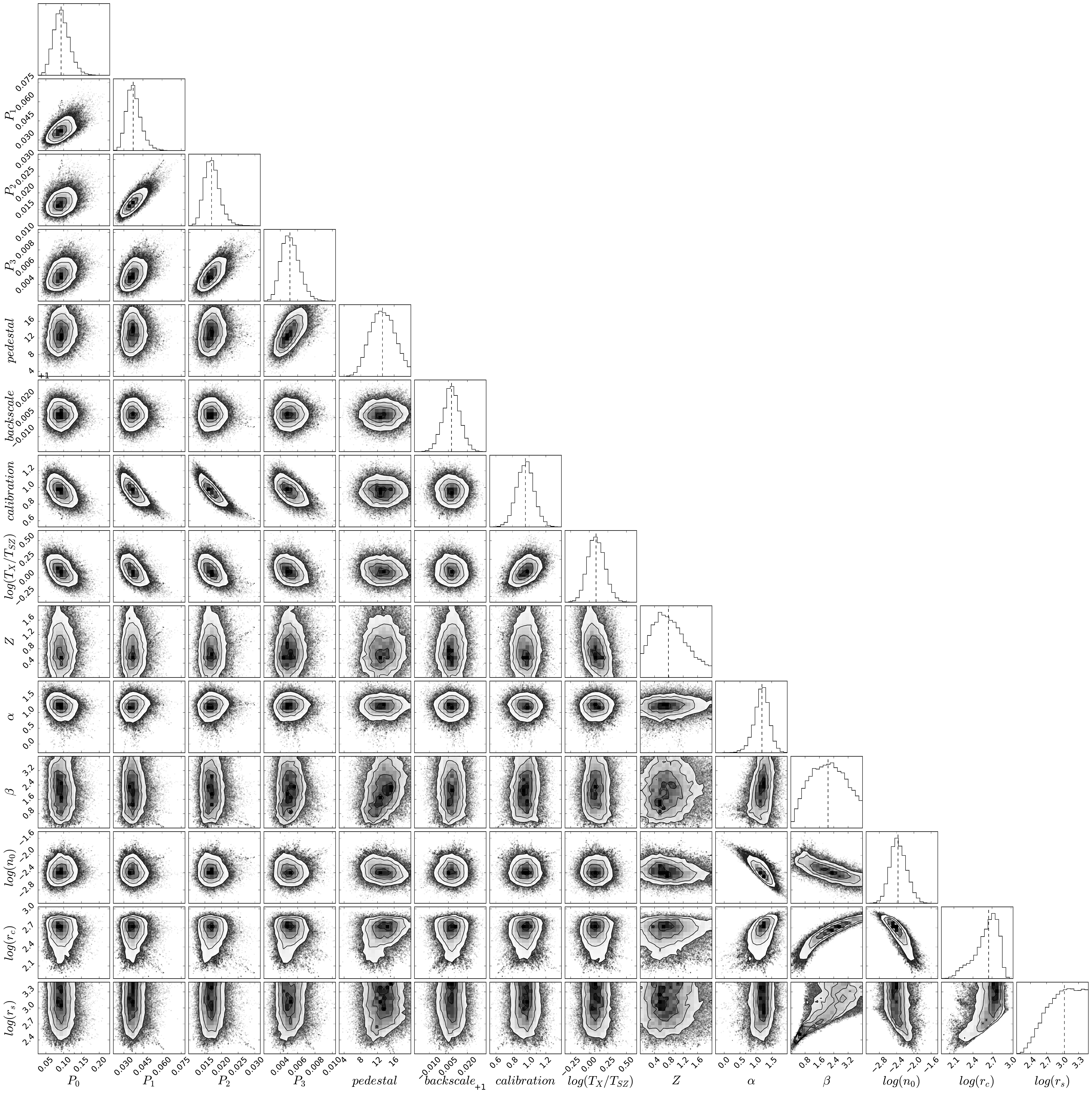}}
\caption[h]{Joint and marginal probability contours for our X+SZ fit. 
Units: keV cm$^{-3}$ for pressure parameters, $\mu K$ for pedestal, Solar for metallicity $Z$, cm$^{-3}$ for density $n_0$, and kpc for radii. As detailed in the paper, uniform priors were taken for all parameters except the calibration, on the same scale ($\log$ when $\log$ are quoted). For the calibration, the Gaussian prior is
very close to the marginal posterior.
}
\label{fig:cornerplot_joint}
\end{figure*}

The model fits the data quite well, in an indistinguishable way from what is
shown for the two separate fits in Figs.~\ref{fig:surf_bright} and \ref{fig:x-ray}. 
All parameters show
well-behaved posterior distributions with little or well-known covariances (Fig.~\ref{fig:cornerplot_joint}).
The relative calibration is very close to 1, $0.95\pm0.10$ 
(uncertainty is determined by the prior), the four $P_i$ values, 
$0.093\pm0.022, 0.037\pm0.005, 0.017\pm 0.002, 0.005\pm0.001$ keV cm$^{-3}$ and pedestal,
$13\pm3$ $\mu$K are consistent with those found in SZ-only fit. As in the X-ray only analysis, 
the backscale continues to be extremely close to 1 ($1.007\pm0.007$) and metallicity is large although with large errors, $Z_{Fe}=0.8\pm0.5$ Solar. The X-ray temperature turns out to be larger
than the SZ temperature $\log(T_X/T_{SZ})=0.09\pm0.11$ because
pressure from SZ data is a bit low compared to the X-ray derived
one within 300 kpc (Fig.~\ref{fig:thermo_profs_x}). Our posterior value is consistent with the known
temperature calibration systematics of Chandra (e.g., Schellenberger et al. 2015). 
We verified that the electron density profile is largely independent of $\epsilon$ by running our code with a wide prior on $\epsilon$. 

\begin{figure}
\centerline{\includegraphics[width=9truecm]{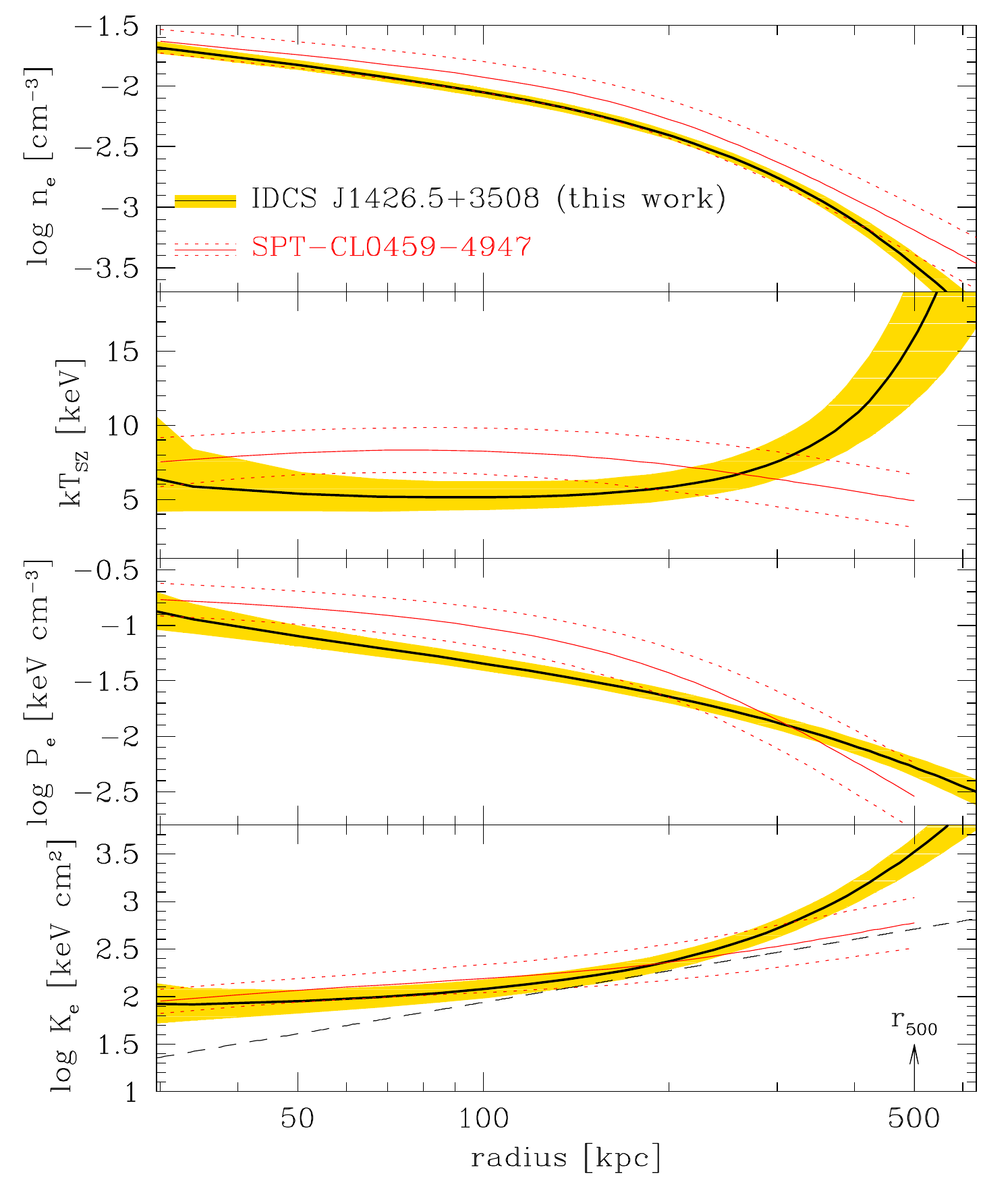}}
\caption[h]{\IDCS thermodynamic profiles 
(mean value and 68\% uncertainties) from the X-ray+SZ analysis (black line and shading).
To illustrate the quality of our determination, the thermodynamic profiles and
approximated $\pm 1\sigma$ errors around the average of SPT-CL0459-4947 are also shown.
SPT-CL0459-4947 is a slightly brighter
and hotter cluster at slightly lower redshift, and its profiles are derived from a 70\%
longer Chandra observations and 500 ks XMM in place of 18 ks MUSTANG-2. 
The dashed black line in the entropy
panel is the Voit et al. (2005) fit to non-radiative simulations, 
as adapted by Pratt et al. (2010). The arrow indicates \IDCS $r_{500}$.}
\label{fig:thermo_profs_joint}
\end{figure}

\begin{figure}
\centerline{\includegraphics[width=9truecm]{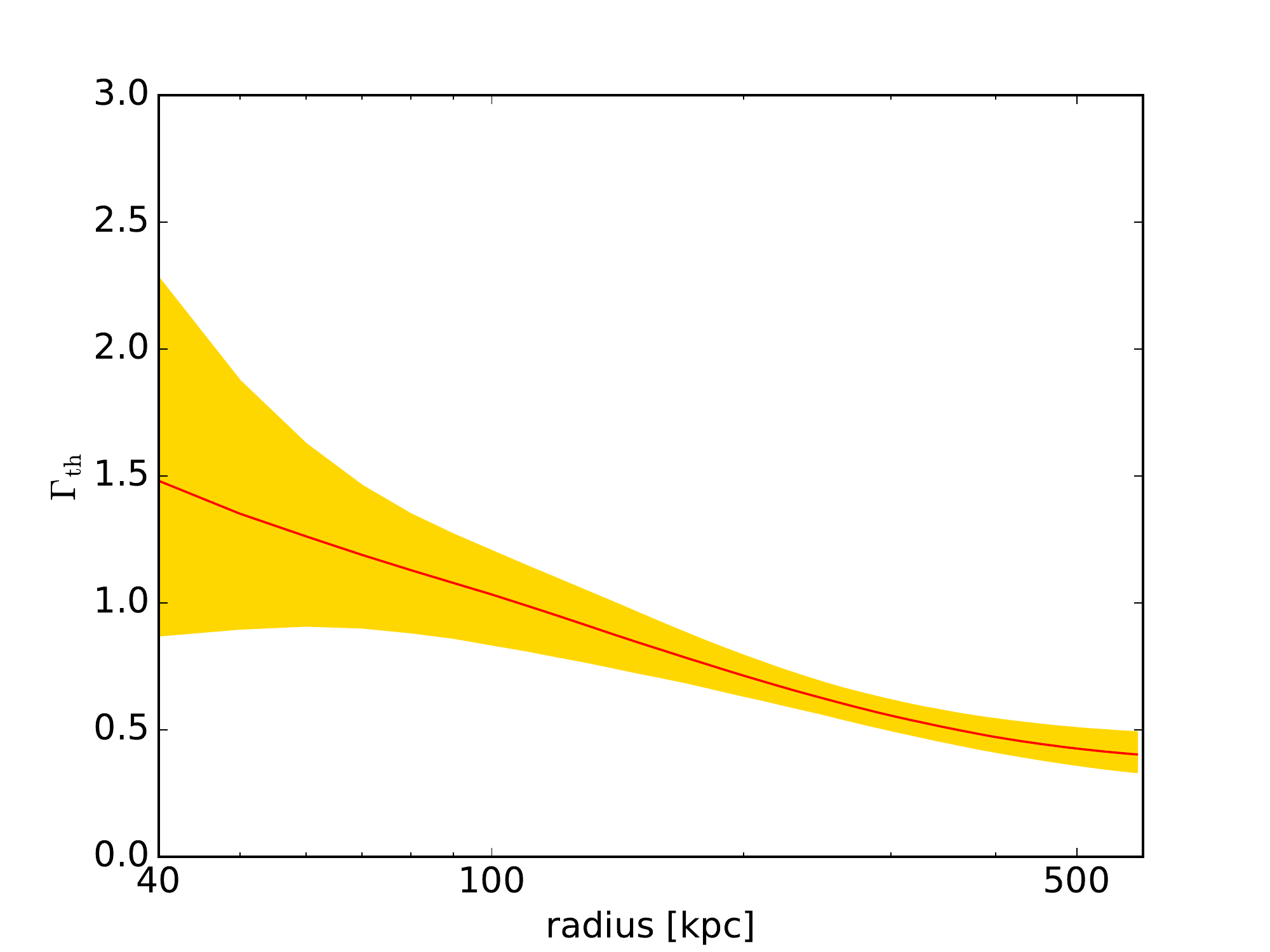}}
\caption[h]{\IDCS thermal polytropic index profile $\Gamma$ 
(mean value and 68\% uncertainties) from the X-ray+SZ analysis (black line and shading). The low value at large clustercentric distances indicate gas motion or turbulence.}
\label{fig:poly_prof}
\end{figure}

Figure~\ref{fig:thermo_profs_joint} shows the derived 
thermodynamic profiles. As expected, the pressure determination improves significantly
when SZ data are included (compared to Fig.~\ref{fig:thermo_profs_x}), and consequently the 
temperature profile also has smaller uncertainties.
Since, as mentioned, at large radii the SZ pressure profile is flatter than the electron density profile (Fig.~\ref{fig:thermo_profs_x}) 
the temperature profile turns up, which is anticipated. 
Because of the outward increasing temperature profile,
the entropy profile turns up, while in most clusters it has a slope equal or lower than
expected from non-radiative simulations (black dashed line). The 
thermal polytropic index, $\Gamma_\mathrm{th}= d \log P/d \log \rho$,  
computed using the pressure derived above which is only sensitive to the thermal component,
turns out to be about $1.2\pm0.2$ at 70 kpc to decrease to $0.47\pm0.06$  at $r \gtrsim 400$ kpc
(Fig.~\ref{fig:poly_prof}).  
Numerical simulations (Borgani et al. 2004; Shaw et al. 2010)
suggest that the polytropic index computed using the total gas pressure, not just the
thermal component, is about 1.2. Assuming that the simulations are correct and that the total pressure follows an adiabatic index of 1.2, the lower observed
values at large radii for the thermal component  indicates an important non-thermal support, i.e., gas motion or turbulence.

The radially increasing temperature, or, equivalently, the presence of 
a larger pressure than for a constant temperature profile, is robustly determined and
unavoidable.
First, the electron density profile at larger radii cannot be made shallower:
given that the total number of collected photons at a fixed radius is fixed, the only way to
flatten the $n_e$ profile is to decrease the background radial profile, leading to a large
increase in the percentage of the net number of cluster photons at large radii. The background however
is constrained by the observed value at large off-axis angles and by the shape measured in
the control fields, so that the normalisation of the 
template background radial profile has almost no freedom (the backscale parameter has less
than 1\% uncertainty, as mentioned). Concerning pressure, the flux density out to 1 arcmin ($\approx 500$ kpc)
is well determined (Fig.~\ref{fig:surf_bright}) and, as a consequence, there is little freedom to force  the 
pressure profile to bend here or at smaller radii, not even applying a strong prior on
its slope. Also, CARMA data (Brodwin et al. 2012) confirm the pressure value at these
radii (Fig.~\ref{fig:pressure}). Since the pressure is well measured at $\approx 500$ kpc, 
changing the outer radius in the Abel (line of sight) integration has a negligible effect on the
recovered pressure profile within 500 kpc. Therefore, a radially increasing temperature
profile is, given the SZ and X-ray data, inescapable. 

An increasing temperature profile over a sizeable part of the cluster is an uncommon feature
in clusters at much lower redshifts, but not a novelty; for example, Abell 2034 (Owers et al. 2014) and  
MACS0417.5-1154 (Botteon et al. 2018) present large outer regions hotter than the 
cluster center and body, as is also true of PLCKG266.6-27.3 at $z=1.0$ (Bartalucci et al. 2018). Again at
low redshift, Abell 2390, which is a plausible descendant of \IDCS (sec.~3.2) also has a monotonically
increasing temperature profile until $r_{500}$ (Vikhlinin et al. 2006).
Equivalently, RXJ1347-1145 (e.g., Kitayama et al. 2004, Mason et al. 2010), and MACS0744 (Korngut et al. 2011) 
present an SZ enhancement outside the core interpreted as a temperature
increase.
If anything, in \IDCS we observe the temperature gradient well into a more 
peripheral region than usually observed in 
nearby clusters, but this could be due
to the difficulties of detecting a temperature gradient in 
the cluster faint outskirts with current X-ray facilities. 

Current \IDCS SZ and X-ray data have insufficient depth to allow 
us to discern in two dimensional maps tiny morphological features, such as
brightness edges, discontinuities or cavities.  This leaves the interpretation
of the increasing temperature profile open: for example, the temperature can be
low at the center because of possible cavities
produced by bubbles of radio plasma emitted by
a central AGN, as proposed by Vikhlinin et al. (2006) to explain the raising 
temperature profile of Abell 2390. Or we may
be faced with an extreme case of sloshing, such as
in Abell 2142 and Perseus (Rossetti et al. 2013, Simionescu et al. 2012). The latter is also corroborated
by the offset of the X-ray emission compared to the BCG location, likewise present in \IDCS.
Our paper shows that by joining X-ray and SZ data, no matter the precise interpretation, disturbances can be effectively detected via
the different slopes of the electron density and pressure profiles (i.e.
via SZ-based temperature profiles).

\section{Discussion}

\subsection{The Efficiency of Combining SZ and X-ray Observations}

Fig.~\ref{fig:thermo_profs_joint} contrasts \IDCS 
profiles to those
of a cluster
at slightly lower redshift, SPT-CL0459-4947 at $z=1.70$ (Ghirardini et al. 2021) derived
by combining a longer Chandra exposure  ($170$ vs $100$ ks)
and a exceptional XMM exposure on an high redshift cluster in place of a typical
MUSTANG-2 observation ($500$ vs $18$ ks)\footnote{Because of the lack of published covariance terms
across parameters of a given profile and across thermodynamic quantities,
plotted errors on SPT-CL0459-4947 
only use the published diagonal terms. Note that the authors use base 10 $\log$ in plots, but we
verified than many quantities quoted with $\log$ use
instead Neperian (natural) $\log$s.}. Prior to our work, SPT-CL0459-4947
was the most distant cluster with well resolved thermodynamic profiles. The 
uncertainty on density and temperature are comparable for the two clusters,
perhaps a bit smaller for \IDCS in spite of the handicap of its lower X-ray 
luminosity
(lower $n_e$ at all radii), while pressure is much better determined for \IDCS.
To summarize, when jointly fitting to a moderate X-ray exposure, a regular MUSTANG-2
observation compares well against exceptional X-ray allocations for the derivation of 
thermodynamic profiles.
Furthermore, our Chandra plus MUSTANG-2 observations of \IDCS favorably compares in terms of precision
to the Ghirardini
et al. (2021) combination of 7 $z_{med}=1.4$ clusters, totaling over 2 Ms Chandra and 
XMM observations.

\begin{figure}
\centerline{\includegraphics[width=9truecm]{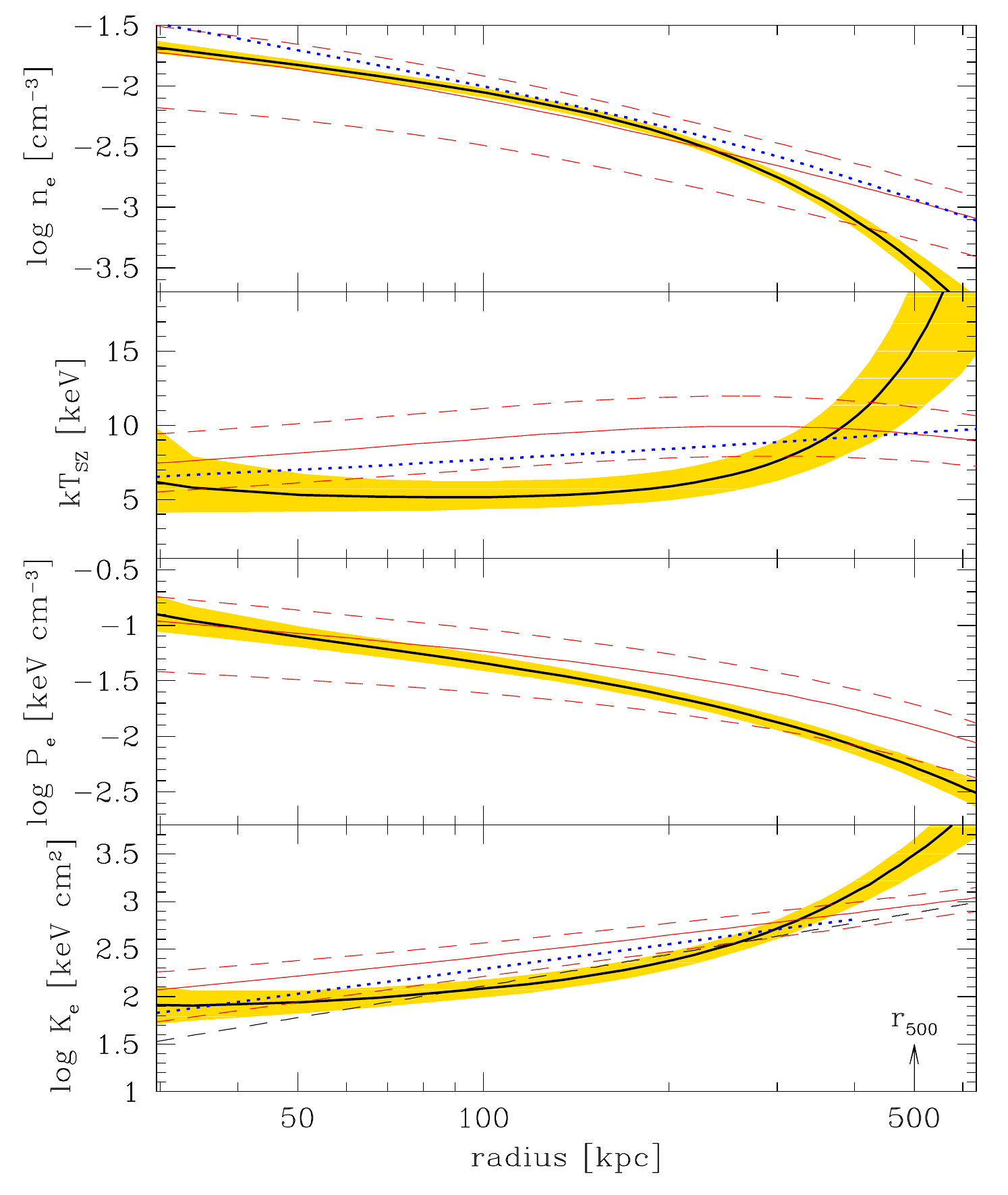}}
\caption[h]{\IDCS 
(mean value and 68\% uncertainties, black line and yellow shading)
is compared to its present-day expected descendant (red line with dashed corridor
mark the mean value and the $\pm2\sigma$, based on X-COP). Compared to its present-day
descendant,
at large radii gas is lacking 
and that present is too hot and with large entropy. To become a present-day
cluster, heat and entropy should be dissipated or the gas transported to larger radii plus cold and low 
entropy gas should be acquired from peripheral regions. In electron density, temperature and entropy panels the
profiles of Abell 2390, a possible \IDCS descendant, are shown with dotted blue line. 
Abell 2390 lacks a published pressure profile measurement.}
\label{fig:compare_descendant}
\end{figure}

\begin{figure}
\centerline{\includegraphics[width=9truecm]{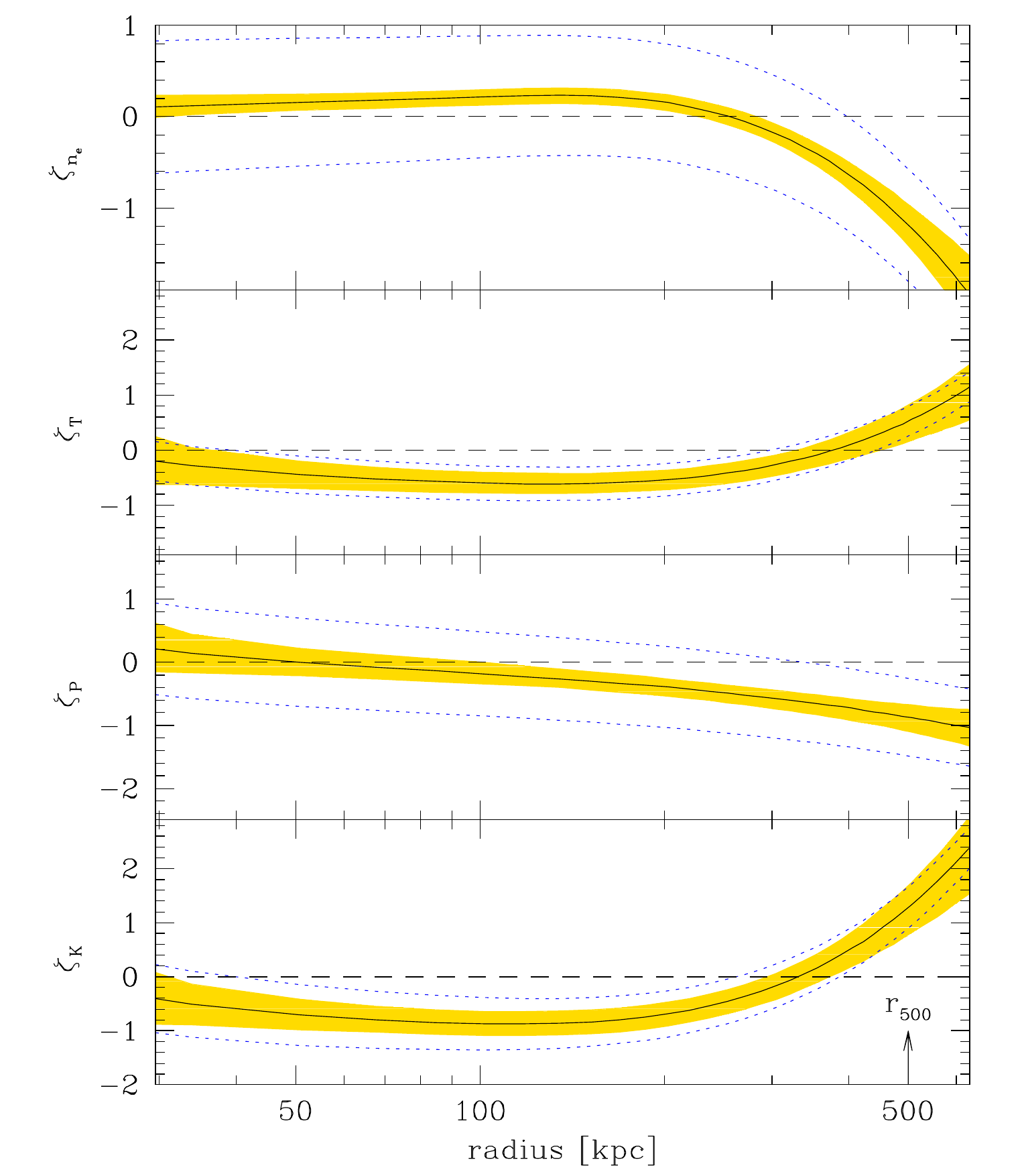}}
\caption[h]{Evolutionary rates $\zeta$ measured at fixed progenitor.
Redshift evolution (slope with $\log E(z)$) of the \IDCS thermodynamic profiles 
(mean value and 68\% uncertainties, black line and yellow shading)
when compared to its present-day expected descendants.
The dotted corridor marks three times the 
X-COP scatter about the average. 
Negative values indicate deficits in the past, while the no-evolutionary 
case correspond to $\zeta=0$. The plot quantifies the evolutionary rates already
appreciable in Fig.~\ref{fig:compare_descendant}. 
}
\label{fig:evolution}
\end{figure}

\subsection{Evolution}

\IDCS is one of the most massive clusters at high redshift (Brodwin et al. 2012).
By selecting clusters of the same mass in the closest
snapshot ($z=1.71$) and their $z=0$ descendants in  Multidark Planck 2 simulation
(Klypin et al. 2016, Behroozi et al. 2013), we find that
by $z=0$ the mass of \IDCS will plausibly have
growth by $0.65\pm0.12$ dex. 
Therefore, we compare \IDCS with present-day
$0.65$ dex more massive clusters. We verified that the $0.12$ dex scatter on the mass growth is largely
subdominant for our conclusions compared to the variety of observed profiles at $z\sim0$ at fixed mass.
We use unscaled radii and unscaled thermodynamic quantities, see Sec.~3.3 for discussion.

Figure~\ref{fig:compare_descendant} compares the thermodynamic profiles of \IDCS with its 
$z=0.07$ plausible descendants (solid red line, the corridor encloses $\pm 2\sigma$, based on X-COP, Ghirardini et al. 2019). 
Since X-COP clusters have lower masses than \IDCS descendants, their profiles are scaled up based on X-COP mass-dependent relations determined
from a scaling measured at lower masses. To test this extrapolation, we also plot electron 
density, temperature, and entropy
profiles (from Vikhlinin et al. 2006 and Sonkamble et al. 2015) of Abell 2390, whose
mass matches the one \IDCS will have by $z=0.23$ (the Abell 2390 redshift). Abell 2390 is well within
the X-COP $\pm 2\sigma$ bounds, confirming the correctness of our extrapolation. 

The comparison of \IDCS to its plausible descendant identifies three radial regimes showing different
behaviours. At the very center, $r=30$ kpc, the gas had the same
thermodynamic values 10 Gyr ago as today. If interpreted as AGN feedback regulating
the inner cluster region (McDonald et al. 2017), this implies a fine tuning between AGN activity 
and gas cooling, to keep all four thermodynamic quantities close to constant with time in the last 10 Gyr.

Outside the center and within 300 kpc, the gas amount in \IDCS is
within the $2\sigma$ range of comparable low-redshift clusters.
However, at about 500 kpc (about $r_{500}$), the electron density profile shows a drop and 
\IDCS is gas-poor relative to present-day descendants, which is understandable because
descendants have accreted mass at large radii (see also Sec.~3.3). 
Within 300 kpc, the gas already present in \IDCS  
is colder and has lower entropy
than in massive clusters at $z=0$, while the opposite is true around 500 kpc.
Therefore, to become a present-day cluster, heat and entropy must
be transferred inwards from larger radii. At the same time a large gas deposit must
occur at 500 kpc   
with little net gas transfer at smaller
radii given the already appropriate electron density profile there. 
Therefore, outside the center and within 300 kpc the emerging picture is one in which the gas 
of \IDCS is already present with the right amount but still far, in terms of
temperature and entropy, from the properties of present-day clusters. 
Instead,
the region at about 500 kpc was not yet developed 10 Gyr ago and the gas present there 
must either move away or evacuate the excess heat and entropy. The low thermal polytropic index supports
this interpretation.
This is theoretically feasible by a filamentary gas stream (Zinger et al. 2016), which can
bring low entropy gas to $r_{500}$ and bring energy into central regions which dissipates in form of heat,
and shown from the hydrodynamic simulations in Sec.~3.3.

Figure~\ref{fig:evolution} quantifies the evolution of the thermodynamic profiles $f$ by
computing  their evolutionary rates $\zeta$. $\zeta$ is the power of the $E(z)^\zeta$
dependence, 
\begin{equation}
f(r,z, M_z) = f(r,z=0, M_{z=0}) E(z)^\zeta \ ,
\end{equation}
where $H(z) = H_0 E(z)$, and $M_{z=0}$ is the mass of the $z=0$ plausible
descendant of the cluster with mass $M_z$ at redshift $z$. Clusters with different masses at different redshifts 
are considered in the attempt
to emulate tracing a single cluster through time by comparing plausible
ancestors and descendants
(in the Sec.~3.4 we will consider different
cases). 
Using 
log-quantities, $\zeta$ is given by the $\log$ difference 
between profiles at \IDCS redshift $z$ and the reference profiles, 
divided by the baseline $\log E(z_{IDCS})-\log E(z_{X-COP})$. A value of zero would 
therefore mean no $E(z)$ dependence at all.
Fig.~\ref{fig:evolution} shows the derived evolutionary rates and
quantifies trends already appreciable in Fig.~\ref{fig:compare_descendant}:
in short, all thermodynamic quantities show almost no evolution at the center; 
within $\approx300$ kpc, the density remains roughly constant, but heating
and entropy increases are needed; at about 500 kpc, the density must increase
while entropy and temperature decrease.

\begin{figure}
\centerline{\includegraphics[trim={0 0 30 30}, clip,width=9truecm]{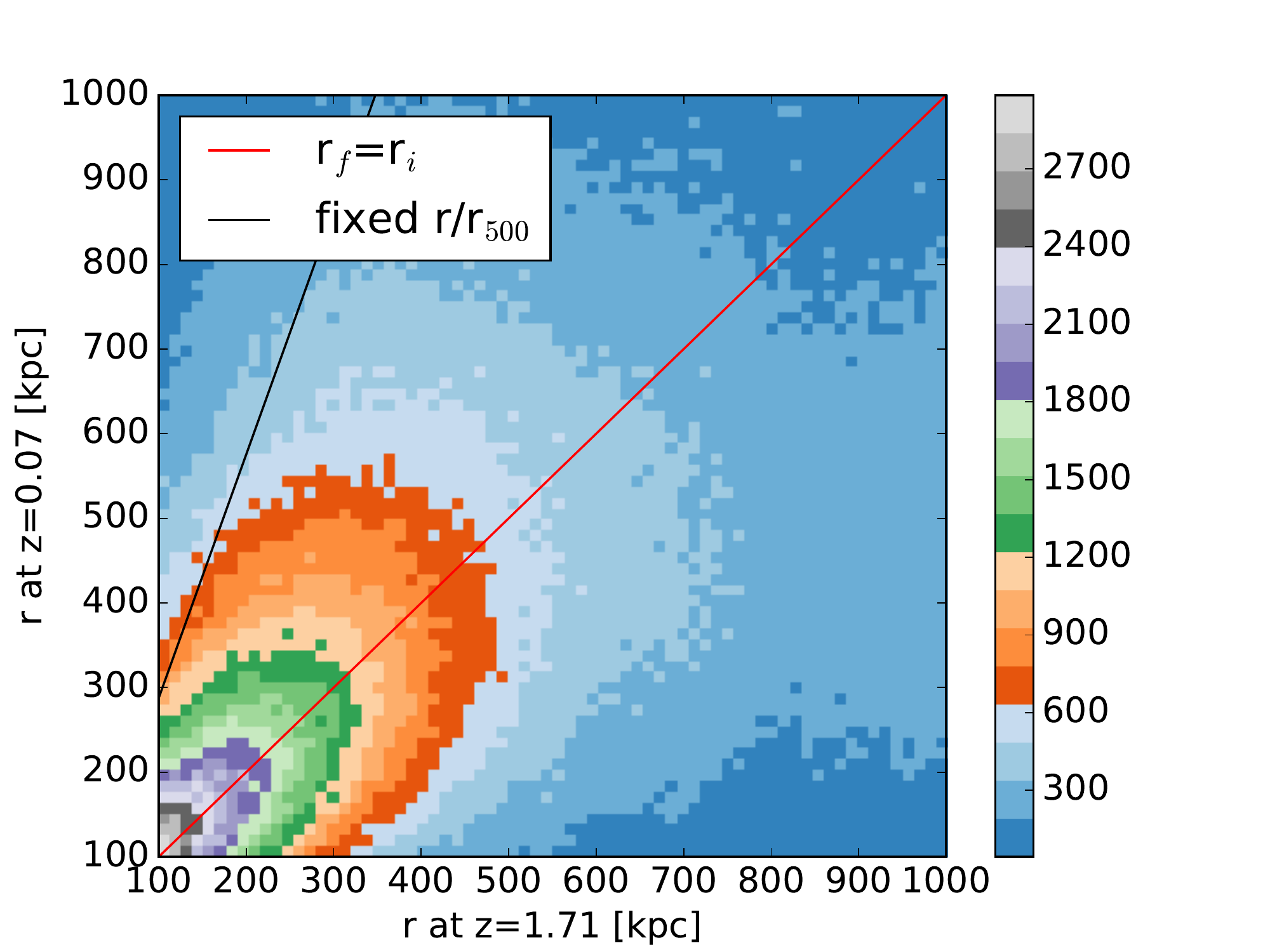}}
\caption[h]{ Two-dimensional histogram showing the distribution of $z=0.07$ clustercentric distances (ordinate) of dark matter particles as a function of their clustercentric distances at $z=1.71$ (abscissa)
in massive clusters of the Magneticum
simulation.  When vertically read, this 2D histogram plot indicates the present-day spatial distribution of the particles that were at distance indicated by their abscissa. The color scale is proportional to their relative frequency.
The locus of constant distance from the cluster
center is shown in red, whereas the black lines show
the locus to be observed if particle distances grow as $r_{500}$, i.e. if $r/r_{500}$ stays constant. Clearly, particles
stay at similar distances from the cluster center in the last 10 Gyr, far less than the factor of three increase
experienced by $r_{500}$. 
}
\label{fig:rinrfin}
\end{figure}

\subsection{Interpretation of results}

We use the Magneticum simulation (Dolag et al. 2016, Ragagnin et al. 2017, Hirschmann et al. 2015) to place our results in a broader context.  The simulation self-consistently follows radiative cooling, star formation, black hole growth, metal enrichment, and associated feedback processes from both Type II/Ia supernovae and AGN. We consider the highest resolution with a 500 comoving Mpc$^3$ volume run (Hirschmann et al. 2015), the 22 most massive clusters at $z=1.71$ and their
$z=0.07$ descendants. In spite of the simulation large volume and our choice of focusing on the most massive clusters, 
the considered halos have similar, yet smaller, masses than \IDCS ($0.64<M_{500} / 10^{14} M_\odot<1.4$ vs 2.5). Nevertheless,
the selected halos are adequate to highlight known theoretical behaviours, as detailed below.
The mass $M_{500}$ of the simulated clusters evolves by $0.60\pm0.16$ dex, and their $r_{500}$ by a factor close to $3$, which is
close to what was found for more massive clusters in the baryon-free, large volume, Multidark simulation in Sec.~3.2.

\begin{figure}
\centerline{\includegraphics[trim={0 0 30 30}, clip,width=9truecm]{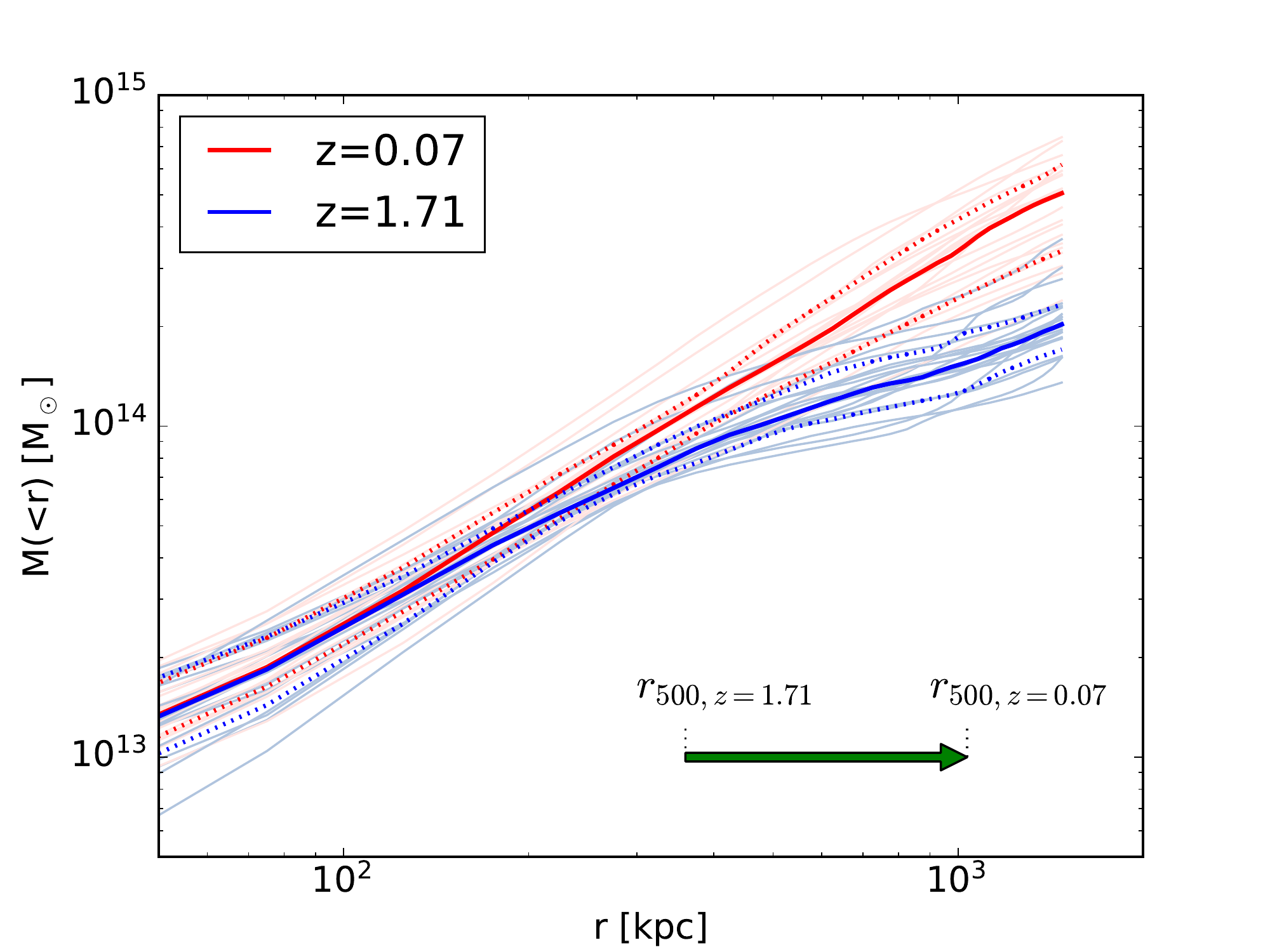}}
\caption[h]{Cumulative mass of massive progenitors and their descendants in the Magneticum simulation.
The solid line indicates the mean profile,
the dotted corridor the 68\% range, individual profiles are plotted in light red or blue.
While $r_{500}$ increases, and by a sizeable amount in the considered redshift interval (shown by the arrow), clusters keep
similar mass density profiles within $r_{500,z=1.71}$ while, at larger radii, particle distances decrease.
}
\label{fig:cumulM}
\end{figure}

Cluster growth is known to proceed in two phases: an early fast accretion, building the cluster central region  largely not evolving
at later times,
and a later phase growing the less central  part of the cluster (e.g. Gott \& Rees 1975, Gunn 1977,
and later works,  e.g. Zhao et al. 2003, Lu et al. 2006). As nicely phrased by B\"{o}hringer et al. (2012), the Birkhoff theorem states that ``the cluster evolution is not concerned with the background universe", clusters do not expand/contract in an expanding/contracting universe, or, to
quote Peebles (1980) ``a gravitational bound system such as the Local group is not expanding"). Indeed, Fig.~\ref{fig:rinrfin} shows that
dark matter particles that were at $r<500$ at high redshift are found at
much the same radius at low $z$,  as already known since Peebles (1980), in agreement with dark-matter only
simulations in Zhao et al. (2003) and with the simple simulation in Lu et al. (2006). The figure also shows  some more accretion from larger radii, in particular
at $r\approx 1000$ (today) kpc, better visible in Fig.~\ref{fig:cumulM},  illustrating the known accretion during the late phase, $z<1.71$ in our case. While $r_{500}$ {\it increases} by a factor of 3, mostly
because it is anchored to the critical density (pseudo-evolution), radial distances of particles
{\it decrease}, or stay constant within $r_{500,z=1.71}$, which is
the only portion of the profile usually accessible at high redshift. Clusters do not expand, and
this explains why
we compare properties at unscaled (i.e. same) radii, which is highly appropriate for observed
portion of \IDCS (within $r_{500,z=1.71}$), unlike many previous
works (e.g.  McDonald et al. 2017, Bartalucci et al. 2017, and Ghirardini et al. 2021)
that preferred instead to compare at fixed  $r/r_{500}$ (shown as black line in Fig.~\ref{fig:rinrfin}), i.e.
consider larger clustercentric distances with decreasing
redshift, while particle distances stay similar or change in the opposite direction.

Mostly because of the pseudo-evolution of $r_{500}$, the mass of the cluster increases, even in absence
of any mass accretion. This implies that
a  comparison at fixed mass is
not comparing ancestors and descendants, but ``unripe apples to ripe oranges in understanding how fruit ripens"
(verbatim from Andreon \& Ettori 1999).  Therefore, our estimates of evolution are derived comparing ancestors and
descendants (at evolving mass for \IDCS ), unlike some works discussed in the next section. 

\begin{figure}
\centerline{\includegraphics[trim={20 0 300 30}, clip,width=7truecm]{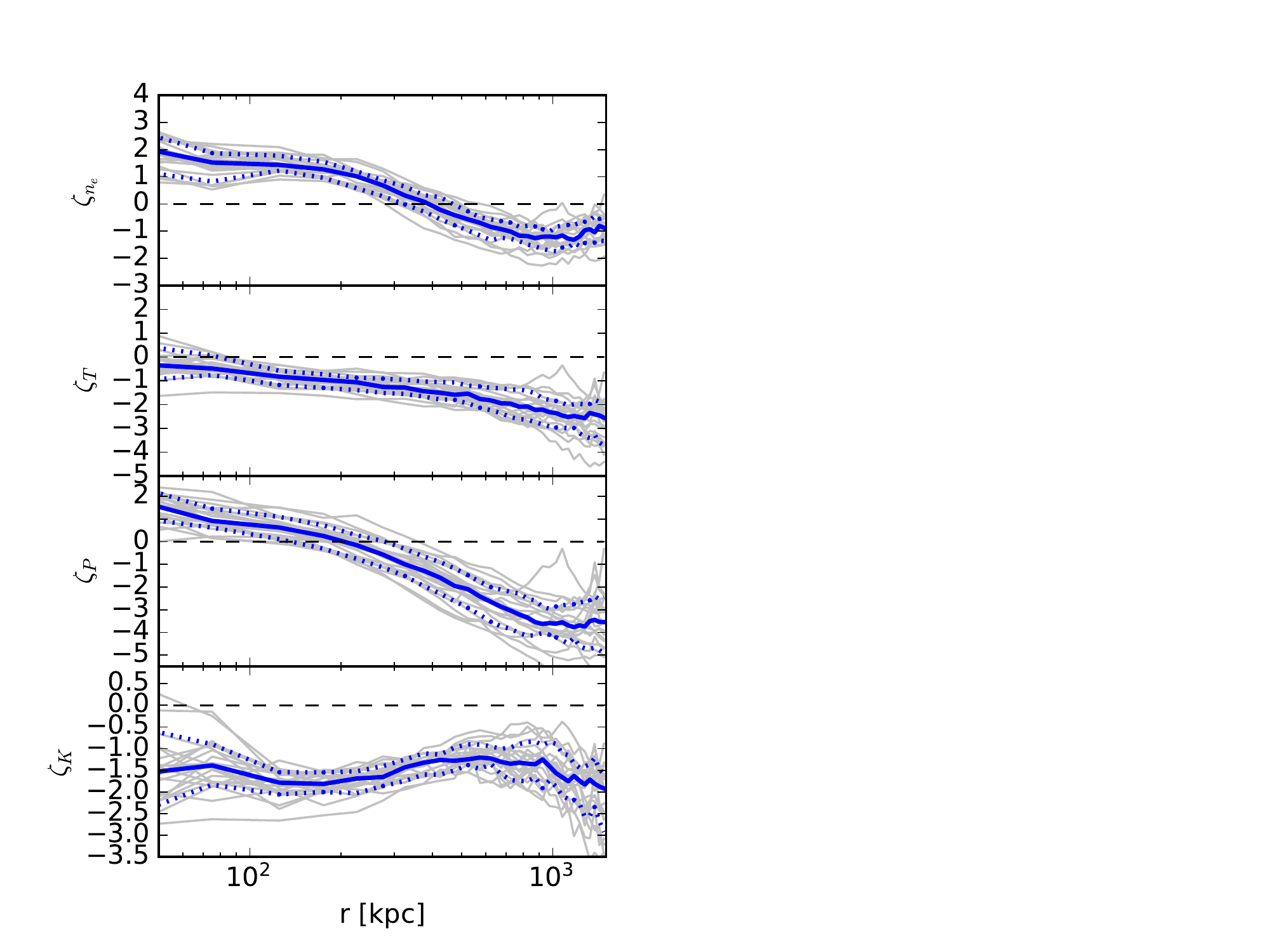}}
\caption[h]{Radial profiles of the evolutionary rates of gas density (upper panel), Temperature (second panel from the top),
Pressure (third panel from the top), and Entropy (bottom panel)
of simulated clusters. The solid line indicates the mean profile,
the dotted corridor the 68\% range, individual profiles are plotted in light gray. The horizontal dashed line marks
the no-evolutionary case. Positive values indicates denser, hotter, higher pressure and entropy at high redshifts.
}
\label{fig:simulunscaled}
\end{figure}

As mentioned, the Magneticum simulation has and follows gas particles.   
Fig.~\ref{fig:simulunscaled} shows the evolutionary rates $\zeta$, derived for gas particles. At high redshift,
clusters were denser (in gas density) in the center (positive $\zeta_{n_e}$) and smaller because the outer regions have not yet grown. They were
also colder (negative $\zeta_{T}$), and had a lower entropy (negative $\zeta_{K}$). Since pressure is given by the product of electron
density and temperature, its evolution is the sum (because of the use of log) of the evolution of the two quantities. 

These trends
are qualitatively similar to what is derived using real data (compare Fig.~\ref{fig:simulunscaled} with Fig.~\ref{fig:evolution})
and what we infer comparing high and low redshift clusters corresponds to what we observe following gas particles in the simulation:
as we wrote for \IDCS in Sec.~3.2, the gas already present in the progenitor within about half $r_{500,z=1.71}$ is colder and
has lower entropy than in the descendant and indeed heat and entropy is
transferred inward, as we concluded for \IDCS. As found with the data, in the last 10 Gyr a large gas deposit occurs at $\approx r_{500,z=1.71}$,
with low net gas transfer at much smaller radii (lower for \IDCS than for simulations). 
In Sec.~3.2 we argued that \IDCS gas at $\approx r_{500,z=1.71}$
must either move away or evacuate the excess heat and entropy, also supported by the low value of the polytropic index at 
$r\gtrsim r_{500,z=1.71}/2$. The bulk motion is obvious following the gas particles in the simulation, as the
temperature and entropy change (Fig.~\ref{fig:simulunscaled}).
There are quantitative differences, however, between simulations and data: $\zeta$ amplitudes tend to be larger in simulation 
and at $r\gtrsim 500$ kpc, where the entropy \IDCS profile shape strongly deviates
from the nearly-constant behaviour seen in the simulated data. In particular, none of the individual simulated
profiles matches, not even qualitatively, the large increase seen for \IDCS . Larger volume simulations, or zoomed-in simulation of more
massive clusters, are needed to understand if differences are due to (imprecise) assumptions on the simulation sub-grid physics or 
differences in the mass of the considered objects.

\subsection{Comparison with literature}

As discussed in the previous section, 
since clusters are not expanding (Peebles 1980, Fig.s~\ref{fig:rinrfin}, \ref{fig:cumulM}) and
with the purpose of comparing apples to apples,
we did the comparison using unscaled units (kpc, keV/cm$^3$, etc) and
we compared \IDCS profiles with those of present-day clusters
$0.65$ dex more massive. 
Given the large redshift baselines presently available, we propose that this choice is
preferable to
comparing clusters with the same mass $M_{500}$.
For example,
Bartalucci et al. (2017) low redshift sample is unlikely to be the descendant 
of their studied $z\sim1$ sample and has actually lower mass if mass were measured inside the same physical aperture.
Comparisons at fixed mass is compounding cluster evolution, dependence of profiles on cluster mass, and pseudo-evolution. The same problem holds using a mass
enclosed in an overdensity anchored
to the Universe matter density, such as with $M_{\Delta,m}$.

\begin{figure}
\centerline{\includegraphics[width=9truecm]{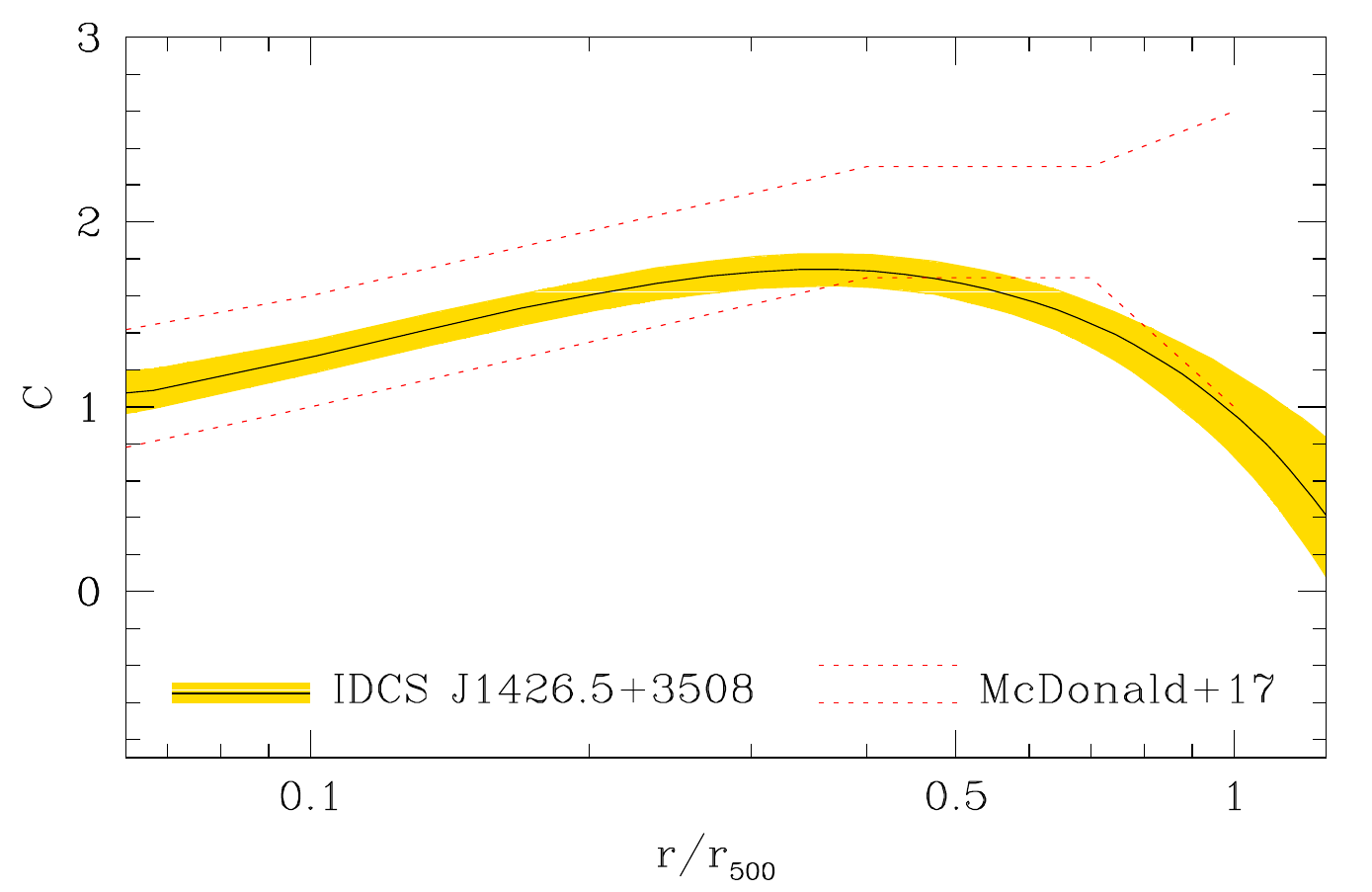}}
\caption[h]{Electron density evolutionary rate 
$C$ computed following McDonald et al. (2017).
The evolution of the \IDCS electron density profile is indicated with the black line and yellow shading
(mean value and 68\% uncertainties), and the McDonald et al. (2017) determination as the $\pm 1\sigma$ dotted corridor.
The low value of $C$ at $r\sim r_{500}$
indicates the need for a large gas density mass growth with this way of calculating the evolutionary rate.}
\label{fig:comp_mcdonald}
\end{figure}

McDonald et al. (2017) computed the evolutionary rate of the electron density only, for
lack of adequate data on other thermodynamic quantities, comparing 
a small $z\sim 1.4$ sample with lower redshift samples whose
average mass increases with decreasing redshift in the attempt of comparing ancestors and descendants, as we also
do.  Similar to
us, they used unscaled electron densities (see inset of their Fig.~3), but different from us, they measured the
power of the $E(z)$ dependency  using scaled radii, unlike our choice of using unscaled radii,
\begin{equation}
n_e(r/r_{500,z},z)= n_e(r/r_{500,z=0},z=0) E(z)^C \ .
\end{equation}

This way of computing evolution
complicates the interpretation of the results because an unique value of rate $C$ takes different
meanings at different radii: for example, 
a cluster with static profiles, such as the simulated halos up to $r\approx r_{500,z=1.71}$, which therefore evolve at the same rate at all radii, would have $C=0$ at the center but $C\ne0$ at $r/r_{500}$ just because different $r_{500}$ would be used at different $z$. Furthermore,
the rate found at some radius $r$ would differ from the one derived at a different radius $r'$.
Therefore, we reiterate, the evolutionary rate derived using scaled radii $C$ compounds evolution and pseudo-evolution.

Fig.~\ref{fig:comp_mcdonald} show the \IDCS electron density evolutionary rate computed following McDonald et al. (2017)
using however X-COP in place of their $z=0$ sample because the latter is not available in the needed form.  
The $C$ value close to zero at large radii indicates 
that the \IDCS profile is dropping at those radii as their descendants do at much larger
radii (at the same scaled radius $r/r_{500}$), not that the profile is unchanging here.
Our and their
electron density evolutionary rates agree (Fig.~\ref{fig:comp_mcdonald}) at $1\sigma$, although 
the authors interpret their uncertain $C$ at large radius as $C=2$, ruled out
for \IDCS given its better S/N.  These
electron density rates complement each other:
our rate uses a high S/N determination of one individual cluster at very high redshift, 
while their rate uses lower signal to noise data and is based on a sample, which 
should average out individual peculiar behaviours. 
Because of data limitations, the authors cannot address temperature, pressure, and entropy
evolution.

Ghirardini et al. (2021) determine the evolutionary rate of seven clusters at $z_{median}=1.4$ mixing the two
approaches above, making their choice completely orthogonal to ours.
First, they used scaled radii, which compounds evolution and
pseudo-evolution. Second, their comparison is at fixed mass, which compounds cluster evolution, 
dependence of profiles on cluster mass, and pseudo-evolution. In addition, the local cluster
sample used
is not representative at the considered (low for the present day) masses used in
their comparison
because their sample is selected against clusters of low pressure,  which
exist in the mass range of their interest, such as CL2015 (Andreon et al. 2019).

The above differences between the ways evolution rates are computed have been overlooked in
the past, which means that evolutionary claims should be reconsidered in
light of the limitations of the used evolutionary rate definition. 
Agreement/disagreement among results of these authors should also be revisited adopting a common
definition of evolutionary rate. Our proposed one, using unscaled profiles and radii at fixed ancestor, has the advantage of separating cluster evolution, dependence on mass, pseudo-evolution
and returns a number whose interpretation is the same at all radii.

\section{Conclusions}

We present well resolved thermodynamic profiles out to 500 kpc (about $r_{500}$) of the $z=1.75$  
galaxy cluster \IDCS. We combined intensity and spectral information from Chandra 
with an SZ map from MUSTANG-2. Thanks to this dataset combination, 
\IDCS becomes the most distant cluster with
resolved profiles, and also the high redshift cluster with 
most precise thermodynamic profiles.  Profiles are distributed as supplementary material.

The profiles are derived using \texttt{JoXSZ} (Castagna \& Andreon 2020)
assuming a non-parametric pressure profile, to allow deviations from the universal pressure profile, 
and a very flexible model for the electron density profile. The analysis accounts
for the 
instrument
calibration uncertainties, background level systematics for both X-ray and SZ data, PSF and
transfer function, and allows the X-ray temperature to differ from the SZ temperature, for
example as a result of the known temperature systematics of Chandra.

The shape of the pressure profile, either derived from the SZ map alone or in conjunction
with the X-ray data, turns out to be flatter than 
the universal pressure profile.  Departures from the universal pressure profile, such those
observed for \IDCS, show the risk of assuming an universal pressure profile when detecting
clusters or measuring integrated pressure (often referred as ``mass''). 
Those may be responsible of the 
discrepancies found 
across instruments that are sensitive 
to different spatial scales.

The \IDCS temperature profile is radially increasing to $r_{500}$ and
the ICM
presents an excess entropy compared to that expected in non-radiative simulations at large radii.
We verified that these behaviours are robustly determined and derive from an 
electron density profile steeper than the pressure profile, both tightly constrained by
the data. 

The \IDCS local plausible descendants will be $0.65\pm0.12$ dex more massive according to our analysis of
Multidark Planck 2 simulations. We therefore infer the evolution of ICM thermodynamics
by comparing \IDCS to their massive descendants using unscaled radii and thermodynamic quantities.
We identify three different radial regimes:
in the very inner region, at 30 kpc, thermodynamic quantities have very similar values
10 Gyrs ago as today. Outside this region and
within 300 kpc (about $r_{500}/2$), the gas is present in the right amount although cooler than in present-day
clusters. Near 500 kpc (about $r_{500}$) \IDCS is short of gas,  the gas present is too hot and has too much
entropy, and the polytropic index is low. Therefore, heat and entropy should
flow away from the region at 500 kpc and into smaller radii. This should happen
with a large gas transfer at about 500 kpc to make the density profile shallower there,
and little net gas transfer at smaller
radii given the already appropriate electron density profile observed there. Non-thermal
pressure, i.e. gas motion or turbulence, 
is also supported by the low polytropic index computed using the data, which
are only sensitive to thermal pressure.
The emerging picture
is therefore one of a cluster in which a filamentary gas stream brings low entropy gas to $r_{500}$
and energy to even smaller radii, which dissipates in form of heat, while ``central" values are
unchanging, indicating a fine tuning between gas cooling and heating there.

To explore the possibly diverse evolution experienced by $z\sim2$ clusters, observations of more  
high-redshift clusters are clearly needed.  JKCS\,041 at $z=1.803$ (Andreon et al. 2009, 2014; Newman et al. 2014) is just being observed and will reported soon.

In this work, we also introduce a new definition  of
the evolutionary rate
which uses unscaled radii and thermodymamic quantities at fixed progenitor,  i.e. which compares clusters of different masses at different redshifts in order to comparing ancestors with descendants. It has the advantage of separating cluster evolution, 
dependence on mass, pseudo-evolution and returns a number with a unique interpretation, 
unlike definitions used in literature. 
Finally, we show that a 100 ks Chandra exposure and a regular MUSTANG-2 observation
compete favorably against a longer Chandra exposure and an XMM observation of exceptional duration
in the determination of thermodynamic profiles.

\section*{Acknowledgements}
F.C. acknowledges financial contribution from the agreement ASI-INAF n.2017-14-H.0 
and PRIN MIUR 2015 Cosmology and Fundamental Physics: Illuminating the Dark Universe with Euclid.
The Green Bank Observatory is a facility of the National Science Foundation operated under cooperative agreement by Associated Universities, Inc. MUSTANG-2, on the GBT, data was taken under the project ID AGBT18A\_336.
The CosmoSim database used in this paper is a service by the Leibniz-Institute for Astrophysics Potsdam (AIP) and SA thanks Harry Enke for suggestions about its use.
The MultiDark database was developed in cooperation with the Spanish MultiDark Consolider Project CSD2009-00064.

\section*{Data Availability}

Raw Chandra data are public available in the Chandra archive, obs ID 15168 and 16321. SZ derived products (beam, transfer function,  and map) are available at https://doi.org/10.7910/DVN/Q1QMQS .
Radial thermodynamic profiles are distributed as supplement material and at the URL above.

{}

\bsp	
\label{lastpage}
\end{document}